\documentclass[12pt]{article}
\usepackage{latexsym}
\usepackage{amsmath}
\usepackage{amssymb}
\usepackage{epsfig,graphics}
\usepackage{graphicx}
\usepackage{booktabs}
\usepackage{multirow}
\usepackage{tikz}
\usepackage{bm}
\usepackage{graphicx}
\usepackage{epsf}
\usepackage{epsfig}
\usepackage{color}
\usepackage{dcolumn}     
\usepackage{amsfonts,amssymb}
\usepackage{dsfont}
\usepackage{slashed}
\usepackage{subfigure}
\usepackage{longtable}

 \usepackage[nottoc]{tocbibind}
\usepackage[colorlinks=true,linkcolor=blue,urlcolor=magenta, citecolor=blue]{hyperref}
\usepackage{fullpage}
\usepackage{amssymb,graphicx}
\usepackage{amsmath}
\usepackage[margin=0.5cm]{caption}
\usepackage{hyperref} 
\usepackage{cite}
\usepackage{upgreek}

\def\l{\left(}
\def\r{\right)}

\newcommand{\be}{\begin{equation}}
\newcommand{\ee}{\end{equation}}
\newcommand{\bea}{\begin{eqnarray}}
\newcommand{\eea}{\end{eqnarray}}
\newcommand{\bg}{\begin{gather}}
\newcommand{\eg}{\end{gather}}
\newcommand{\bseq}{\begin{subequations}}
\newcommand{\eseq}{\end{subequations}}

\renewcommand{\ln}{\mathop{\rm ln}\nolimits}

\newcommand{\Tr}{{\rm Tr}}

\newcommand{\otoprule}{\midrule[\heavyrulewidth]}

\begin{document}

\begin{titlepage}
\begin{center}
{\large\bf Excitations of  Ising Strings on a Lattice}\\
\vspace{0.5cm}
{ \large
Andreas Athenodorou$^{a,b}$, Sergei Dubovsky$^{c}$, Conghuan Luo$^c$, \\ \vspace{0.1cm} and Michael Teper$^{d}$ 
}\\
\vspace{.45cm}
{\small\textit{$^a$ Computation-based Science and Technology Research Center,\\
The Cyprus Institute, Cyprus
}}\\
\vspace{0.3cm}
{\small\textit{$^b$ Dipartimento di Fisica, Universit\'a di Pisa and INFN,\\
Sezione di Pisa, Largo Pontecorvo 3, 
56127 Pisa, Italy
}}\\
\vspace{0.3cm}
{\small  \textit{  $^c$Center for Cosmology and Particle Physics,\\ Department of Physics,
      New York University\\
      New York, NY, 10003, USA}}\\ 
      \vspace{.3cm}  
      {\small \textit {{$^d$Rudolf Peierls Centre for Theoretical
Physics,\\Clarendon Laboratory, University of Oxford,\\ Parks Road, Oxford OX1 3PU, UK\\
\centerline{and}
All Souls College, University of Oxford,\\
High Street, Oxford OX1 4AL, UK}}}

\end{center}
\vspace*{0.4in}
\begin{abstract}
The 3d Ising model in the low temperature (ferromagnetic) phase describes dynamics of two-dimensional surfaces---domain walls between clusters of parallel spins.
The Kramers--Wannier duality maps these surfaces into worldsheets of confining strings in the Wegner's ${\mathbb Z}_2$ gauge theory. We study the excitation spectrum of long Ising strings
by simulating the  ${\mathbb Z}_2$ gauge theory on a lattice. We observe a strong mixing between string excitations and the lightest glueball state and do not find 
indications for light massive resonances on the string worldsheet.
\end{abstract}


\vspace*{0.95in}


\end{titlepage}

\setcounter{page}{1}
\newpage
\pagestyle{plain}

\tableofcontents

\section{Introduction}
\label{section_intro}

The Ising model has been a  fruitful area of research since its discovery in 1920's \cite{ising1925contribution}. 
The 3d Ising universality class is realized in a number of physical systems such as 3d uni-axial magnets \cite{MATTSSON1994L23} and liquid-vapor critical points \cite{Sullivan_2000}. 
On the theoretical side, a lot of work has been devoted over the years to the physics of the 3d Ising model and to calculations of its observables, such as critical exponents.
A celebrated example of a successful approach is provided by the $\epsilon$-expansion  \cite{Wilson:1971dc}.
 Over the last decade, an impressive progress has been achieved by the numerical conformal bootstrap \cite{El-Showk:2012cjh,Kos:2014bka,Kos:2016ysd}, which fixes critical exponents and OPE coefficients of the 3d Ising model to the greatest precision. Monte-Carlo simulations also give very precise results for the critical exponents of the 3d Ising model (see, e.g.,\cite{hasenbusch2010finite}). 
 
 Still this leaves one wondering whether a better analytical control is possible over the 3d Ising model, especially given that the 2d Ising model is exactly solvable. A particularly intriguing set of ideas~\cite{Polyakov:1987ez,Distler:1992rr} is related to the possibility of rewriting the 3d Ising model as a theory of (super)strings. In this description the string worldsheet corresponds to a boundary between clusters of positive and negative spins. In the 2d Ising model the corresponding boundaries describe worldlines of free Majorana particles, which gives rise to an expectation for fermionic excitations to be present on the string worldsheet in the 3d case.  This idea has been realized explicitly in the lattice phase of the Ising model \cite{dotsenko1988fermion}, however, the continuum description of the Ising strings is still missing. The corresponding string theory is expected to be strongly coupled, however see \cite{Iqbal:2020msy} for an interesting recent proposal towards a weakly coupled description.
 
 Given this state of affairs it is natural to explore the structure of the Ising strings experimentally, where by experiments we mean  lattice Monte--Carlo simulations.  
 For this purpose it is convenient to use the 3d version of the Kramers--Wannier duality, which maps the low energy ferromagnetic phase of the 3d Ising model into a confining phase of the  $\mathbb{Z}_2$ lattice gauge theory \cite{Wegner:1971app}.
 Under this duality, Ising domain walls are mapped  into worldsheets of $\mathbb{Z}_2$ confining strings\footnote{Actually this map is more subtle. When we approach criticality, there will be proliferation of domain walls with high genera\cite{Caselle:1993yu,Caselle:1994aka}. The worldsheet described by effective string theory is better considered as a macroscopic effective description of the condensate of these genuine domain walls. We thank Michele Caselle for pointing this out. }. 
 To gain insight into the worldsheet dynamics it is natural to focus on the so-called long strings (or torelons). These are strings wrapped around one of the compact spatial dimensions.
  The ground state energy and the first few lowest-lying states of Ising strings 
 in the long string sector have been previously studied in \cite{Caselle:2002ah,Caselle:2005vq,Caselle:2006dv}. 
%
%
%
%

In this work, we aim to extend these results with a more precise spectrum calculation and to determine energies of a larger number of excited states. 
Excitations of closed flux tubes wrapped around one of the spatial dimensions are characterized by their longitudinal momentum $q$ along the flux tube. In addition,
 one may  also define two parity transformations. The longitudinal parity $P_l$ corresponds to a reflection along the string and maps $q$ to $-q$. The transverse parity $P_t$ corresponds to a reflection in the transverse direction. The main goal of our study is to check whether Ising strings carry massive resonant states on their worldsheet. 
 Our initial results seem to indicate the presence of a massive resonance  in the parity $(++)$ sector (at $q=0$). The same state is also present at the lowest non-vanishing $q$\footnote{Recall that the values of $q$ are quantized as a result of a compactification on a circle.}. However, a careful analysis shows that this state is a bulk glueball rather than a new worldsheet state. 

Similar string spectrum computations were previously performed  in the 3d  $U(1)$ gauge theory \cite{Athenodorou:2018sab} and in the 3d and 4d  $SU(N)$ Yang-Mills theories \cite{Athenodorou:2010cs,Athenodorou:2011rx,Athenodorou:2013ioa,Athenodorou:2021vkw}. In these studies, massive resonances are  observed in some cases, such as for the fundamental 4d $SU(N)$  confining string and confining strings in higher representations.
Quite surprisingly though, fundamental confining strings in 3d $SU(N)$ gluodynamics don't show any sign of additional massive resonant modes on the string worldsheet. 

We see that Ising strings are in some sense in between these two options. On one side, we observe a well-pronounced resonant state in the spectrum of torelon excitations. On the other hand, this is not a new state, but rather a bulk glueball. This  strong mixing between torelon excitations and glueballs is possible due to the absence of large $N$ suppression in the Ising case.  In this case, the effective string theory makes sense below the threshold corresponding to the emission of glueballs. 

%

The rest of the paper is organized as follows.
In section \ref{section_ising}, we review  properties of the 3d Ising model and  its duality to the $\mathbb{Z}_2$ lattice gauge theory. 
 In section \ref{section_effective_string} we review the basics of the effective string theory, which provides a good approximation for the lowest-lying spectrum. In section \ref{section_lattice} we summarize the basics of the lattice gauge theory and of the Monte-Carlo simulations. We describe  the algorithm for computing the closed flux tube spectrum, and discuss how we reduce the systematic and statistical errors and improve the projection onto low-lying states. 
 In section \ref{section_results} we present our results for some of the basic parameters such as the string tension and the lightest glueball mass. We present and analyze the closed flux tube spectra in 3d $\mathbb{Z}_2$ gauge theory for a wide range of string lengths. We start with the absolute ground state and continue onto excited states in different sectors. In particular, we identify a massive resonance state that is not described by the Nambu-Goto theory. Then
  we describe the checks which we performed, which indicate that the observed state is not in fact a novel worldsheet state but rather a scattering state of a long string with an additional unbound glueball. In section \ref{section_conclusions}, we present our conclusions and discuss future directions.

\section{Ising Model and $\mathbb{Z}_2$ Gauge Theory}
\label{section_ising}
The 3d Ising model is one of the simplest spin models of (anti-)ferromagnetism. Its partition function is given by 
\begin{equation}
    Z = \sum_{s_i} e^{-H(s_i)\over T} \,,
\end{equation}
where the Ising Hamiltonian is given by
\begin{equation}
    H(s_i) = -J \sum_{\langle i,j\rangle} s_i s_{j}-h\sum_i s_i\;.
\end{equation}
Here the first sum runs over all neighboring pairs of spins $s_i=\pm1$ on a cubic lattice. In the present paper we are interested in the Ising model with a vanishing external magnetic field
\[
h=0\,.
\]
Then the theory enjoys a global $\mathbb{Z}_2$ symmetry, which flips signs of all spins. Positive values of the coupling constant $J$ correspond to ferromagnetism and negative ones to anti-ferromagnetism. Indeed,
 for positive $J$ the Hamiltonian is smaller for spins pointing in the same direction making it energetically favorable for spins to be aligned. On the other hand, thermal fluctuations tend to randomize the spins. Which effect wins depends on the temperature, so the model exhibits a (second order) phase transition at a critical temperature $T_c$. 
As a consequence of the bipartite property of the square lattice the ferromagnetic and anti-ferromagnetic models are equivalent at $h=0$.
Namely, they can be mapped into each other by taking  $J \rightarrow -J$ and flipping half of the spins, which correspond to one of the sublattices. In what follows we assume 
\[
J>0\;.
\]
At a critical temperature $T=T_c$ the spins develop long range correlations which are described by a conformal field theory.  At temperatures below the critical one
 the global $\mathbb{Z}_2$ symmetry is spontaneously broken and a typical spin configuration describes clusters of positive and negative spins separated by domain walls of positive tension. In the vicinity of the critical temperature,
 \[
 T\lesssim T_c
 \]
 this phase is described by a continuous gapped Ising field theory.
%
%
%
%
%
%
%
As reviewed in the introduction, it is a longstanding question whether it is possible to rewrite the Ising dynamics as a tractable continuum string theory, where the string worldsheet describes the dynamics of the domain walls.
Our goal here is to study the structure of the Ising strings through the lattice Monte-Carlo simulation.

To study the string dynamics it is instructive to map the Ising model into a  $\mathbb{Z}_2$ gauge theory. This map has been constructed by Wegner~\cite{Wegner:1971app} and can be considered as a generalization of the 
Kramers--Wannier duality of the 2d Ising model  (see, e.g., \cite{Savit:1979ny} for a review). Unlike in the 2d Ising model which is self-dual, the duality maps the 3d Ising model into a different theory defined by the following
 partition function
\begin{equation}
\label{Zdual}
    Z_{gauge}(\beta) = \sum_{\{\sigma_l = \pm 1 \}} \exp \left(\beta \sum_{\square} \sigma_{\square} \right)\;.
\end{equation}
Here $\sigma_l$ variables define a $\mathbb{Z}_2$ gauge connection which lives on the links of the dual lattice. The coupling constant $\beta$ of the dual theory is related to the Ising model parameters via
\begin{equation}
    \beta = -\frac{1}{2} \log\mbox{\rm tanh}\, {J\over T}\;.
\end{equation}
This Abelian gauge theory exhibits a number of properties characteristic of the non-Abelian $SU(N)$ Yang--Mills theory. First, it enjoys a global 1-form ${\mathbb Z}_2$ center symmetry (see \cite{Gaiotto:2014kfa} for a modern introduction). Similarly to the $SU(N)$ case, upon compactification on a circle the ${\mathbb Z}_2$ center symmetry is realized by (pseudo)gauge transformations with twisted boundary conditions. A Polyakov loop operator, defined as a Wilson loop wound around the circle, carries a negative ${\mathbb Z}_2$ charge. As a result, in the phase with unbroken center symmetry a sector with a Polyakov loop insertion is orthogonal to a trivial sector with no operators wound around the circle. Analogously to the $SU(N)$ case we will refer to the states created by topologically trivial operators as glueballs. Deformed Polyakov loops acting on a vacuum produce ``long" flux tube states, which are the main target of our study. 

The phase with unbroken center symmetry, which describes the confined phase of the ${\mathbb Z}_2$ gauge theory, is realized at \cite{blote1999cluster}
\[
\beta<\beta_c\approx 0.7614133(22)\;,
\] 
where the critical value $\beta=\beta_c$ corresponds to the conformal Ising point.
In addition,  Ising strings exhibit a roughening transition at  \cite{hasenbusch1997computing}
\[
\beta=\beta_r=0.47542(1)\;,
\]
so we are interested in the range $\beta_r<\beta<\beta_c$, where the string dynamics is described by a continuum theory in the scaling limit $\beta\to \beta_c$.

The deconfining phase transition at $\beta=\beta_c$ needs to be distinguished from the one that happens when the circumference $R $ of the spatial circle gets sufficiently small, namely at \cite{Caselle:2002ah}
\be
\label{RcIsing}
R=R_c\approx{0.82\ell_s}\;,
\ee
where $\ell_s^{-2}$ is the tension of a confining string. 
Later we are going to refer to $l_s$ as the (intrinsic) string width, given that we only have one mass scale in the theory. 
The latter transition corresponds to the finite temperature deconfining phase transition of the ${\mathbb Z}_2$ gauge theory understood as a $(2+1)$-dimensional quantum field theory. The parameter $\beta$ is a coupling constant of this theory,
which also has an interpretation as the inverse temperature, if one understands the ${\mathbb Z}_2$ gauge theory as a 3-dimensional classical statistical model. The Polyakov loop plays a role of the order parameter for both phase transitions.

In principle, both Ising and $\mathbb{Z}_2$ descriptions can be used for Monte-Carlo studies of Ising strings (see, e.g., \cite{mon1988new,Caselle:2002ah,Caselle:2005vq,Caselle:2006dv} for some previous work). In the Ising description this is achieved by introducing ``interfaces", {\it i.e.}, by flipping the sign of the coupling $J$ on the links which intersect the string worldsheet. To study the spectrum of string excitations, which is our main goal here, the gauge theory description appears more convenient. Indeed, in this description excited strings states are created by deformed Polyakov loops. As reviewed  in section~\ref{section_lattice} this makes it straightforward to produce a large basis of excited states by changing the shape of the Polyakov loop. Furthermore, a precision mass determination requires a good overlap of the operator basis with the low lying string states. The gauge theory formulation allows this to be achieved by the well-developed techniques of blocking and smearing.

For future reference, note that in addition to the string tension $\ell_s^{-2}$, the  $Z_2$ gauge theory in the confining phase has another characteristic energy scale---the inverse correlation length $\xi^{-1}$, which is set by the
lightest glueball mass. Given that the parity invariant Ising model has a single relevant deformation, in the scaling limit the ratio of the two scales is universal.
Its numerical value is~\cite{Agostini:1996xy} 
\begin{equation}
\label{xitol}
    {\xi^2\over\ell_s^2} \approx 0.1056(19)\;.
\end{equation}

%
%
%
%
%
\section{Effective String Theory}
\label{section_effective_string}
In the absence of additional symmetries confining strings are not expected to carry any massless states on the worldsheet apart from the $(D-2)$ gapless translational Goldstone bosons describing transverse oscillations of a string.
Here $D$ is the total number of space-time dimensions. In particular, one expects to find a single massless mode on the worldsheet of $D=3$ Ising strings. Then the spectrum of low lying long string excitations is strongly constrained by the non-linearly realized target space Poincar\'e symmetry and can be calculated using the effective string theory (see, e.g.,  \cite{Dubovsky:2012sh,Aharony:2013ipa} for a review). Effective string theory provides a natural reference point to be compared with the actual string spectrum, so let us briefly summarize properties of the effective string spectrum.

The most straightforward approach for calculating the effective string theory predictions is based on the perturbative expansion which uses the ratio ${\ell_s/R}$ as a small parameter.
As a consequence of the non-linearly realized Poincar\'e symmetry all  terms in this expansion up to (and including) ${\cal O}(1/R^5)$ are universal.
This means that those terms are insensitive to the microscopic theory as soon as no additional massless degrees of freedom are present on the worldsheet. This universality provides a powerful self-consistency check for lattice results. On the other hand it makes it quite challenging to probe the underlying  microscopic theory by high precision measurements of
the string ground state for which the  ${\ell_s/R}$ expansion has good convergence properties.

Furthermore, the ${\ell_s/R}$ expansion exhibits poor convergence for excited string states. An efficient technique to calculate the effective string theory predictions for these states is based on the Thermodynamic Bethe Ansatz \cite{Dubovsky:2013gi,Dubovsky:2014fma}, which can also be reformulated as an undressing method based on the $T\bar{T}$ deformation \cite{Chen:2018keo}. In this approach one calculates perturbatively the worldsheet $S$-matrix, and then makes use of a non-perturbative relation between the $S$-matrix and 
the finite volume spectrum to predict the latter. This technique is a close cousin of the familiar L\"uscher method \cite{Luscher:1990ux} combined with the TBA method \cite{Zamolodchikov:1989cf} for calculating the leading order winding corrections, which is possible due to an approximate integrability of the effective string theory. The leading order TBA string spectrum is given by
\begin{equation} \label{nambu_goto}
    E_{GGRT}(N_l, {N_r}) = \sqrt{\frac{4\pi^2(N_l-N_r)^2}{R^2} + { R^2\over \ell_s^4} + {4\pi\over \ell_s^2} \left(N_l+N_r-\frac{D-2}{12} \right)},
\end{equation}
which is nothing but the Goddard--Goldstone--Rebbi--Thorne (GGRT) spectrum \cite{Goddard:1973qh} of a bosonic string in a winding sector.  Here $N_l$ and $N_r$ are non-negative integers called levels, which count the total left- and right-moving momenta along the string.
The total longitudinal momentum is given by 
\begin{equation}
    p= \frac{2\pi (N_l-N_r)}{R}\,.
\end{equation}

In what follows it will be instructive to compare the Ising string spectrum with the GGRT one.
Note that at $D=26$ the GGRT spectum (\ref{nambu_goto}) coincides with the exact spectrum of critical bosonic strings. At $D\neq 3,26$ this spectrum is not compatible with the $D$-dimensional Poincar\'e symmetry and should be considered as a leading order approximation in the $l_s/R$ expansion. The $D=3$ case is somewhat special, and an integrable theory of a single massless boson with the spectrum given by (\ref{nambu_goto}) appears to be a consistent candidate for the worldsheet theory of a long $D=3$ string. Motivated by the lattice data, the confining string of $D=3$ Yang--Mills theory was conjectured to describe a single massless bosons, however, the corresponding spectrum deviates from the $D=3$ GGRT formula. As we will see, for the Ising string the deviations are even more pronounced. 
%
%
%
%
%

The GGRT states are completely characterized by the occupation numbers $n_{l}(k)$, $n_{r}(k)$,  where $k$ is a positive integer labeling  longitudinal momenta. These string excitations are generated by creation operators $a_k$ and $a_{-k}$\footnote{For convenience we omit the $\dagger$. Because the annihilation operators will not appear in this paper, it should cause no confusion. }. We will denote the corresponding state as $ |n_l(k),n_r(k)\rangle$, which is a shorthand notation of $ |n_l(k),n_r(k); \; k=1,2,\dots \rangle$.
Given such a state its levels can be computed as 
\begin{equation}
    N_l = \sum_k n_l(k) k, \quad N_r = \sum_k n_r(k) k\;.
\end{equation}
In what follows we will refer to effective string excitations as phonons. For instance, the $N=\tilde{N}=2$ GGRT level corresponds to two degenerate states. One of these states is a two-phonon excitation with
\[
n_l(2)=n_r(2)=1\;,
\]
and another a four-phonon excitation with
\[
n_l(1)=n_r(1)=2\;,
\]
where in both cases all other phonon occupation numbers vanish.

As discussed in the Introduction, the long string spectrum is invariant under longitudinal and transverse parity transformations $P_l$ and $P_t$.
It is straightforward to determine the corresponding transformation properties of the GGRT states. Namely, as far as the transverse parity is concerned, its action depends
only on the total number of excitations and all GGRT state are eigenvalues of $P_t$,
\be
\label{Pt}
P_t|n_l(k),n_r(k)\rangle=(-1)^{ \sum_k (n_l(k)+n_r(k))}|n_l(k),n_r(k)\rangle\;.
\ee
On the other hand, the longitudinal parity acts by exchanging the left- and right-moving excitations,
\be
\label{P_l}
P_l|n_l(k),n_r(k)\rangle=|n_r(k),n_l(k)\rangle\;.
\ee

Finally, note that in our discussion of the GGRT spectrum we implicitly set the total transverse momentum $p_t$ to zero. By restoring the $p_t$ dependence we obtain the full set of the GGRT states $|n_l(k),n_r(k),p_t\rangle$, with the energies given by the conventional relativistic formula,
\[
E(p_t)=\sqrt{p_t^2+E(0)^2}\;.
\] 
 
For convenience in Table~\ref{table_NGstates} we present the states created by phonon creation operators in different sectors with $q=0,1$ and $N_l+N_r \le 6$. We will discuss more about the quantum numbers that define the sectors in section~\ref{subsection_fluxtube}. 

\begin{table}[htp]
\begin{center}
\centering{\scalebox{0.9}{
\begin{tabular}{c||c||c} \otoprule \otoprule 
\multicolumn{3}{c}{$q=0$} \\
\otoprule \otoprule 
\ \ \ \ \ \ \ \ \ \ \ \ \ \ \ \ $N_l,N_r$ \ \ \ \ \ \ \ \ \ \ \ \ \ \ \ \ &  \ \ \ \ $P_t,P_r$ \ \ \ \  & \ \ \ \ \ \ \ \ \ \ GGRT States \ \ \ \ \ \  \ \ \ \ \\ \otoprule  
 $N_l=N_r=0$ & $++$ & $| 0 \rangle$ \\ \midrule 
$N_l=N_r=1$ &  $++$ & $a_{1} a_{-1} | 0 \rangle$ \\ \midrule 
\multirow{4}*{$N_l=N_r=2$} & $++$ & $a_{2} a_{-2} | 0 \rangle$ \\
&  $++$ & $a_{1} a_{1} a_{-1} a_{-1} | 0 \rangle$ \\
&  $-+$ & $(a_{2}a_{-1} a_{-1} + a_{1}a_{1} a_{-2}) | 0 \rangle$ \\
&  $--$ & $(a_{2}a_{-1} a_{-1} - a_{1}a_{1} a_{-2})  | 0 \rangle$
\\\midrule
\multirow{9}*{$N_l=N_r=3$} & $++$ & $a_{3} a_{-3} | 0 \rangle$ \\
&  $++$ & $a_{2} a_{1} a_{-2} a_{-1} | 0 \rangle$ \\
&  $++$ & $a_{1} a_{1} a_{1} a_{-1} a_{-1} a_{-1} | 0 \rangle$ \\
&  $++$ & $(a_{1} a_{1} a_{1} a_{-3} + a_{3} a_{-1} a_{-1} a_{-1}) | 0 \rangle$ \\
&  $+-$ & $(a_{1} a_{1} a_{1} a_{-3} - a_{3} a_{-1} a_{-1} a_{-1}) | 0 \rangle$ \\
&  $-+$ & $(a_{3} a_{-2} a_{-1} + a_{2} a_{1} a_{-3}) | 0 \rangle$ \\
&  $--$ & $(a_{3} a_{-2} a_{-1} - a_{2} a_{1} a_{-3}) | 0 \rangle$ \\
&  $-+$ & $(a_{2} a_{1} a_{-1}  a_{-1}  a_{-1} + a_{1} a_{1} a_{1} a_{-2} a_{-1}) | 0 \rangle$ \\
&  $--$ & $(a_{2} a_{1} a_{-1}  a_{-1}  a_{-1} - a_{1} a_{1} a_{1} a_{-2} a_{-1}) | 0 \rangle$ \\\otoprule \otoprule 
\multicolumn{3}{c}{$q=1$} \\
\otoprule \otoprule 
\ \ \ \ \ \ \ \ \ \ \ \ \ \ \ \ $N_l,N_r$ \ \ \ \ \ \ \ \ \ \ \ \ \ \ \ \ &  \ \ \ \ $P_t$ \ \ \ \  & \ \ \ \ \ \ \ \ \ \ GGRT States \ \ \ \ \ \  \ \ \ \ \\ 
\otoprule 
 $N_l=1,N_r=0$ &  $-$ & $a_{1}| 0 \rangle$ \\ \midrule 
 \multirow{2}*{$N_l=2,N_r=1$} &  $+$ & $a_{2} a_{-1}| 0 \rangle$ \\
&  $-$ & $a_{1} a_{1} a_{-1} | 0 \rangle$ \\ \midrule
\multirow{6}*{$N_l=3,N_r=2$} &  $+$ & $a_{3} a_{-2} | 0 \rangle$ \\
&  $+$ & $a_{2} a_{1} a_{-1} a_{-1}| 0 \rangle$ \\
&  $+$ & $a_{1} a_{1} a_{1} a_{-2} | 0 \rangle$ \\
&  $-$ & $a_{3}a_{-1} a_{-1} | 0 \rangle$ \\
&  $-$ & $a_{2} a_{1} a_{-2}| 0 \rangle$ \\
&  $-$ & $a_{1} a_{1} a_{1} a_{-1} a_{-1}| 0 \rangle$  
\end{tabular}}}
\end{center}
\caption{\label{table_NGstates}
Table with the states of the lowest GGRT levels with $q=0,1$ and $N_l+N_r \le 6$.}
\end{table}

\section{Review of Lattice Techniques}
\label{section_lattice}

\subsection{Lattice gauge theory and Monte-Carlo simulations}
\label{subsection_lgt}

A general lattice gauge theory (LGT) is described by a set of fields associated with the links of a lattice. 
Lattice links may be labeled by a pair $(n,\mu)$, where $n$ labels the lattice site, and $\mu$ is a direction.
Each lattice link  is then mapped to an element $U_\mu(n)$ of the gauge group. 
For a cubic lattice the action of a LGT is given by
\begin{equation} \label{lattice gauge theory}
    S = \beta \sum_{\rm plaq} \{1- \operatorname{Re} (\Tr \,U_{\rm plaq})\}\,,
\end{equation}
where the sum is over elementary squares (``plaquettes") of the lattice which may be labeled as $(n,\mu,\nu)$ and
\[
U_{\rm plaq}(n,\mu,\nu) = U_{\mu}(n) \cdot U_{\nu}(n+\hat{\mu}) \cdot U_{\mu}^{\dagger}(n+\hat{\nu}) \cdot U_{\nu}^{\dagger}(n)\,,
\]
 is an ordered product of gauge fields around a plaquette. In our case of $\mathbb{Z}_2$ gauge theory, the link elements are simply $\pm 1$ and the action is simplified as:
 \begin{equation}
     S = - \beta \sum_{\rm plaq} U_{\rm plaq} \,.
     \label{Z2 gauge theory}
 \end{equation}
 
 The action (\ref{Z2 gauge theory}) is gauge invariant and can be used to generate Monte-Carlo simulations. Periodic lattices are used in this work. In principle, one can generate millions of configurations using Markov Chain Monte-Carlo algorithms. After achieving thermalization, we compute statistical quantities through importance sampling. Different algorithms may have different thermalization speeds and different step sizes between configurations. In this paper we only use the Metropolis algorithm. For each lattice system, we created 200000 configurations to perform measurements, with 25 sweeps between two measurements in order to reduce auto-correlation. 

Statistical quantities calculated in this work are correlation functions
\begin{equation}
    \langle \phi_i(U) \phi_j(U) \cdots \rangle = \int \prod dU \phi_i(U) \phi_j(U) \cdots e^{-S}\,,
\end{equation}
of gauge invariant operators $\phi_i(U)$'s.
Two-point correlators calculated at different times can be used to extract the spectrum of different physical states such as glueballs and flux tubes. The corresponding procedure is further discussed in section \ref{subsection_extracting_spectra}. 

The lattice spacing $a$ has units of length, but in numerical simulations we only deal with numbers, so we have to choose units where everything is dimensionless. A common choice is to use lattice units, which sets $a=1$. This choice is implicitly assumed in the action expression (\ref{lattice gauge theory}). This choice is convenient during the simulations, but the cost is that the continuum limit becomes obscure. So it is also common to express physical observables using the units defined by a certain characteristic energy scale of interest. In this work we are mostly interested in confining strings, so we will use string units which set the string tension to one, $\ell_s = 1$. 

Independently of the units, the continuum limit is achieved when
\begin{equation}
    {a^2\over \ell_s^2} \rightarrow 0\;.
\end{equation}
Of course, in practice this is impossible to achieve on a finite lattice. At the fixed lattice size the quality of the continuum limit is controlled by the difference between the $\mathbb{Z}_2$ coupling constant $\beta$ and its critical value $\beta_c = 0.7614133(22)$. In order to stay in the confined phase we need to keep $\beta<\beta_c$. Note that we cannot take the difference $\beta-\beta_c$ too small, because otherwise one should use very large lattices in order to make sure that the string width does not exceed the size of the lattice. 

%
%
%
%
\subsection{Extracting spectra}
\label{subsection_extracting_spectra}

In this work we use the framework of \cite{Teper:1998te,Athenodorou:2010cs,Athenodorou:2013ioa} to measure the spectrum. 
Namely we construct a set of operators $\phi_i$ in a sector characterized by certain quantum numbers and acting on constant time slices\footnote{Of course, we work on an Euclidean lattice, so a choice of the ``time" direction is a matter of convention.}. Then a two-point correlator of two operators separated by $n_t$ lattice units in the time direction, which corresponds to the physical time $t=an_t$, can be written in the following form 
\begin{equation}
    C_{ij}(t) = \langle \phi_i^{\dagger} (t) \phi_j (0) \rangle = \sum_k \langle v|\phi_i^{\dagger} e^{-Ht} |k \rangle \langle k| \phi_j|v \rangle = \sum_k c_{ik} c_{kj}^* e^{-E_kt},
\end{equation}
where the sum goes over a complete set  $|k\rangle$ of energy eigenstates with the chosen quantum numbers, $| v\rangle$ is the absolute vacuum state and $c_{ik}$'s are the overlap coefficients
 \[
 c_{ik} = \langle v| \phi_i^{\dagger}| k \rangle\;.
 \]
%

As the time separation increases, higher energy contributions decay faster and only lowest energy states survive. It can be shown \cite{Luscher:1990ck} that at large times the eigenvalues $\lambda_a(t)$  of the matrix $C^{-1}(0)C(t)$ are given by the spectrum,
\begin{equation}
    \lambda_a(t) \approx e^{-tE_a}, \quad t \rightarrow \infty \,,
\end{equation}
if the basis of operators is large enough. To determine the energies in practice one first constructs the approximate eigenstates $\Phi_i$ by diagonalizing the correlation matrix $C^{-1}(0)C(t=a)$ at early times, and then extracts the corresponding energy eigenvalues from the exponential falloff of the diagonal correlation functions $\langle\Phi^\dagger_i(t)\Phi_i(0)\rangle$.
To illustrate this procedure, let us consider the simplest case of a single operator, which allows to determine the ground state energy in the corresponding sector. In this case the diagonalization is trivial, so one simply studies  the correlator 
\begin{equation} 
    \langle \phi^{\dagger}(t) \phi(0) \rangle = \sum_n |\langle v |\phi| n\rangle|^2 e^{-E_n t} \underset{t \rightarrow \infty}{\rightarrow} |\langle v |\phi| 0\rangle|^2 e^{-E_0 t}. 
    \label{diagonal_correlator}
\end{equation}
To analyze its behavior it is convenient to define an effective mass
\begin{equation}
    am_{eff} (t) = -\ln \left( \frac{\langle \phi^{\dagger}(t) \phi (0) \rangle}{\langle \phi^{\dagger}(t-a) \phi (0) \rangle} \right) \,.
    \label{effective_mass}
\end{equation}
In the limit of an infinite statistics it decreases monotonically  over time and asymptotes to the actual ground state energy in the $\phi$ sector,
\begin{equation}
    am_{eff}(t) \underset{t \to \infty}{\to} aE_0. 
\end{equation}
In practice one plots the effective mass as a function of time and extracts $E_0$ from the position of a plateau, which is followed by statistical fluctuations.
 For the ground state, the effective mass sets an upper bound on the actual energy and it is possible to observe the plateau up to rather late times. 
 
 A general strategy for measuring energies of excited states is similar, but the practicalities become more and more challenging for highly excited states.
 Indeed, statistical noise in the measured effective mass is an unavoidable feature of the Monte-Carlo simulations using the importance sampling to compute correlators.
 The amplitude of the noise stays constant in time, while correlators exhibit an exponential decay. Inevitably, at large enough time $t_n$ statistical noise becomes larger than the signal and the effective mass needs to be measured before this happens. Correlators corresponding to heavier excited states decay faster, so that the critical time $t_n$  is shorter for them. 
 
 Clearly, this implies that one needs to achieve a maximal possible overlap of the approximate eigenstates $\Phi_i$ with the true energy eigenstates, so that the plateau can be measured as early as possible.
 On the other hand, given that we perform a diagonalization in an artificially truncated finite dimensional Hilbert space, every approximate eigenstate necessarily has an admixture of heavier states which needs to decay before the plateau can be observed. This problem becomes more and more severe for highly excited states. 

%
%
To overcome this problem one needs to maximize the projection of an approximate eigenstate on the true energy eigenstates. This projection can be estimated by the gap between the value of the effective mass at $t=a$ and the plateau.
Typically, for us this projection drops below $\sim0.5$ around level $N_l,N_r=3$, so we do not expect the corresponding energy determinations to be reliable.

There are several ways to improve a quality of the plateau. 
First, one may try to minimize the measured energies in lattice units. 
This can be achieved by choosing the values of the parameters such that the string tension is smaller in the lattice units. In the Ising model this can be achieved by picking the value of $\beta$ close to the critical point. 
However, other issues arise as one approaches  the critical point. First, as one does this, one needs to take a larger lattice to model a system of the same physical size ({\it i.e.}, as measured in string units). Given that 
we work on a three-dimensional lattice, the simulation time grows as a cube of the lattice size. 
 Also,  close to the critical point, correlations between gauge field configurations created by the Metropolis algorithm become higher. To overcome this one needs to increase the
 sampling interval, which also results in a longer simulation time. All in all, a limited computing power prevents one from approaching the critical point too closely. 

The second way to reduce statistical errors is by creating a larger size of samples. This is also limited by the computing resource. 

Finally,  one can improve the quality of the operators, so that the overlaps of the approximate eigenstates to the exact ones are closer to unity. This can be achieved both by starting with a larger set of operators, and also by suppressing the overlap of the operators with the highly energetic microscopic states using blocking and smearing techniques. 
 We will discuss this more in section \ref{subsection_fluxtube}. 


%
%
\subsection{Constructing flux tube operators}
\label{subsection_fluxtube}

In this paper we work in the confining phase of the $\mathbb{Z}_2$ gauge theory.
Equivalently, this is the phase with an unbroken center symmetry. Recall that given a gauge theory compactified 
on a circle, the center symmetry may be defined\footnote{A modern definition of the center symmetry as a 1-form symmetry does not require to consider a compactification~\cite{Gaiotto:2014kfa} (see also~\cite{Wang:2018edf} for a more recent work). A traditional and less general discussion presented here is enough for our purposes.} by making use of the ``twisted gauge transformation" generated by gauge functions $g$ satisfying
\begin{equation}
\label{twist}
    g(R) = \Lambda g(0)\;,
\end{equation}
where $\Lambda$ is a center element of the gauge group, and $R$ is the circumference of the circle. The Yang-Mills action functional is invariant under such a transformation. 
However, given that the gauge function (\ref{twist}) is not periodic, this transformation defines a global (rather than a gauge) symmetry of the theory.
On the other hand, any two transformations satisfying (\ref{twist}) with the same $\Lambda$ can be related to each other by a conventional gauge transformations. Hence, after dividing out over the conventional gauge transformations,
one obtains a global symmetry transformation which is isomorphic to the center subgroup of the gauge group.
 For $SU(N)$ gauge theory it is the ${\mathbb Z}_N$ center symmetry, and $\Lambda = e^{\frac{2\pi i k}{N}}$. For the ${\mathbb Z}_2$ gauge theory the center symmetry is ${\mathbb Z}_2$ itself. 

This definition makes it clear that an arbitrary Wilson loop
\begin{equation}
    W_C = \Tr \left[ \mathcal{P} \exp(i\oint_C A^{\mu}(x)dx_{\mu}) \right],
\end{equation}
corresponding to the contour $C$ with a trivial winding along the chosen compact direction is neutral under the center symmetry. Indeed, such a loop necessarily crosses any transverse slice an equal number of times from both sides
and all factors of $\Lambda$ cancel out. On the other hand, a Polyakov loop is wound around the periodic dimension, so it crosses any transverse slice in one direction one time more than in the opposite direction. As a result, it is charged under the center symmetry transformation. This also shows that its vacuum expectation value(vev) plays a role of the order parameter for the center symmetry. In the confining phase Polyakov loops have zero vev, and a long string sector is generated by acting on the vacuum by (an arbitrarily deformed) Polyakov loop. Of course, in addition one may add also any number of topologically trivial Wilson loops creating additional glueball states. The center symmetry ensures that this sector does not mix with the topologically trivial one, which is generated by the glueball operators only.

Before describing the set of operators which we used to probe long strings, let us describe conserved quantum numbers in these sector. 
First, there is a longitudinal momentum $p$ along the flux tube. Flux tubes are wound  around a circle of a circumference $R$, so the longitudinal momentum is quantized 
\[
p = \frac{2\pi q}{R}\;,
\] 
with $q$ being an integer. The ground state is translationally invariant, which corresponds to $q=0$. 

In addition, there are two parity transformations $P_t$ and $P_l$, which we already introduced in our discussion of the GGRT spectrum in section~\ref{section_effective_string}. It is straightforward to describe how they act on the gauge theory operators, without any reference to effective strings. Let us consider a long string winding around the $x$ direction. Then the transverse parity is a mirror transformation acting on the transverse $y$ direction,
\[
 (x, y) \xrightarrow{P_t} (x, -y)\;.
 \]
Similarly, the longitudinal parity $P_l$ acts as a mirror transformation of the longitudinal $x$-direction,
\[
 (x, y) \xrightarrow{P_l} (-x, y)\;.
 \]
Note that in general the longitudinal parity does not commute with the longitudinal momentum,
\[
P_l\,p\,P_l=-p\;,
\]
so that only $q=0$ states may be simultaneous eigenstates of $p$ and $P_l$.

Finally, long string states may also carry a non-vanishing transverse momentum $p_t$.
It does not convey any useful information about the worldsheet dynamics and we will always set it to zero by averaging over transverse positions of all operators.   
%

Let us describe now the set of operators, which we use to probe the long string sector. The simplest operator charged under the center symmetry associated to the compact $x$ direction is the straight Polyakov loop
\begin{equation}
\label{strPol}
    \phi_P(y, t) = \prod_{n=1}^{R/a} U_x(x+na, y, t)\;.
\end{equation}
where $R=La$ is the string length. In principle, this operator can be used to measure the ground state energy of a long flux tube. However, its overlap with the ground state of the flux tube is quite poor.
Indeed, the Polyakov loop (\ref{strPol}) creates a string with a width of order the lattice spacing $a$. On the other hand, a physical string close to its ground state is expected to have width of order the characteristic string scale $\ell_s$.

\begin{figure}[t!]
\includegraphics[width=0.6\textwidth]{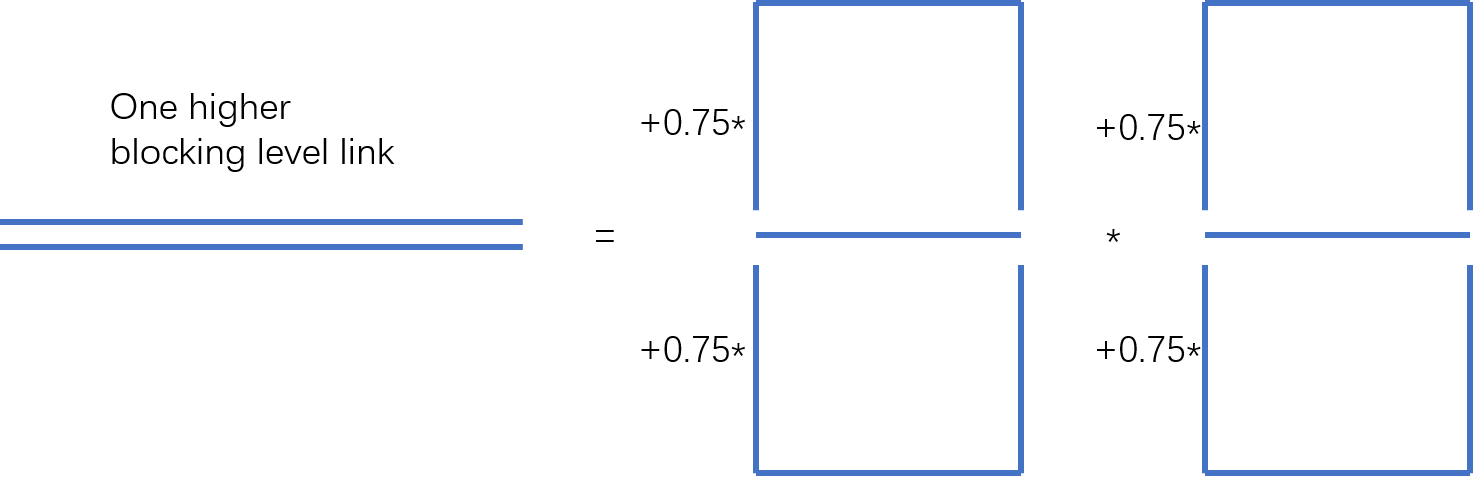}
\centering
\caption{Increasing the blocking level of a link by one.}
\label{blocking}
\end{figure}

The overlap can be improved by applying a combination of smearing and blocking procedures \cite{Teper:1987wt}.
 One starts with the usual link field, which corresponds to blocking level $N_{\rm bl}=1$.
 Then one replaces an original link with a sign of a weighted average over the link itself and two staples attached to it (see Fig. \ref{blocking}). In our simulations we chose the averaging weight to be 0.75\footnote{The weight will in general affect the overlaps onto low-lying states. But the difference is checked to be small for different weights in the $U(1)$ calculation\cite{athenodorou2019spectrum}. Because we have already got good overlaps, we don't try to optimize this parameter in this work. }. Finally,
  one constructs a twice longer link by multiplying two consecutive smeared links. The result is what one calls a level 2 blocked link. To construct the links at $N_{\rm bl}$-th blocking level one applies the same procedure using the
  blocking level $N_{\rm bl}-1$ links as an input.

%
%

\begin{figure}[h!]
\includegraphics[width=0.6\textwidth]{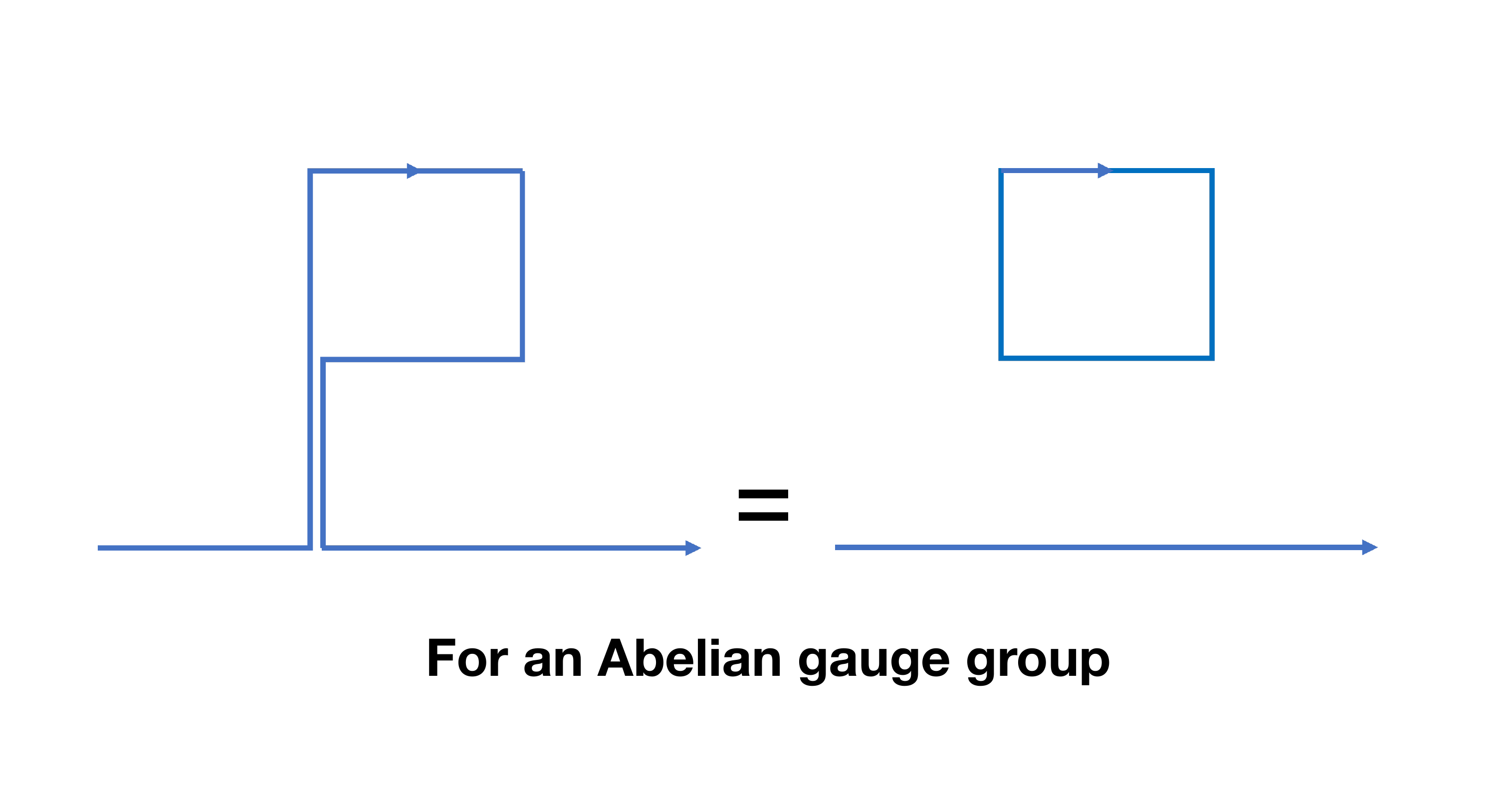}
\centering
\caption{For an Abelian gauge group there is no sharp distinction between string excitations and additional glueballs.}
\label{multitrace}
\end{figure}

\begin{figure}[t!]
\includegraphics[width=0.7\textwidth]{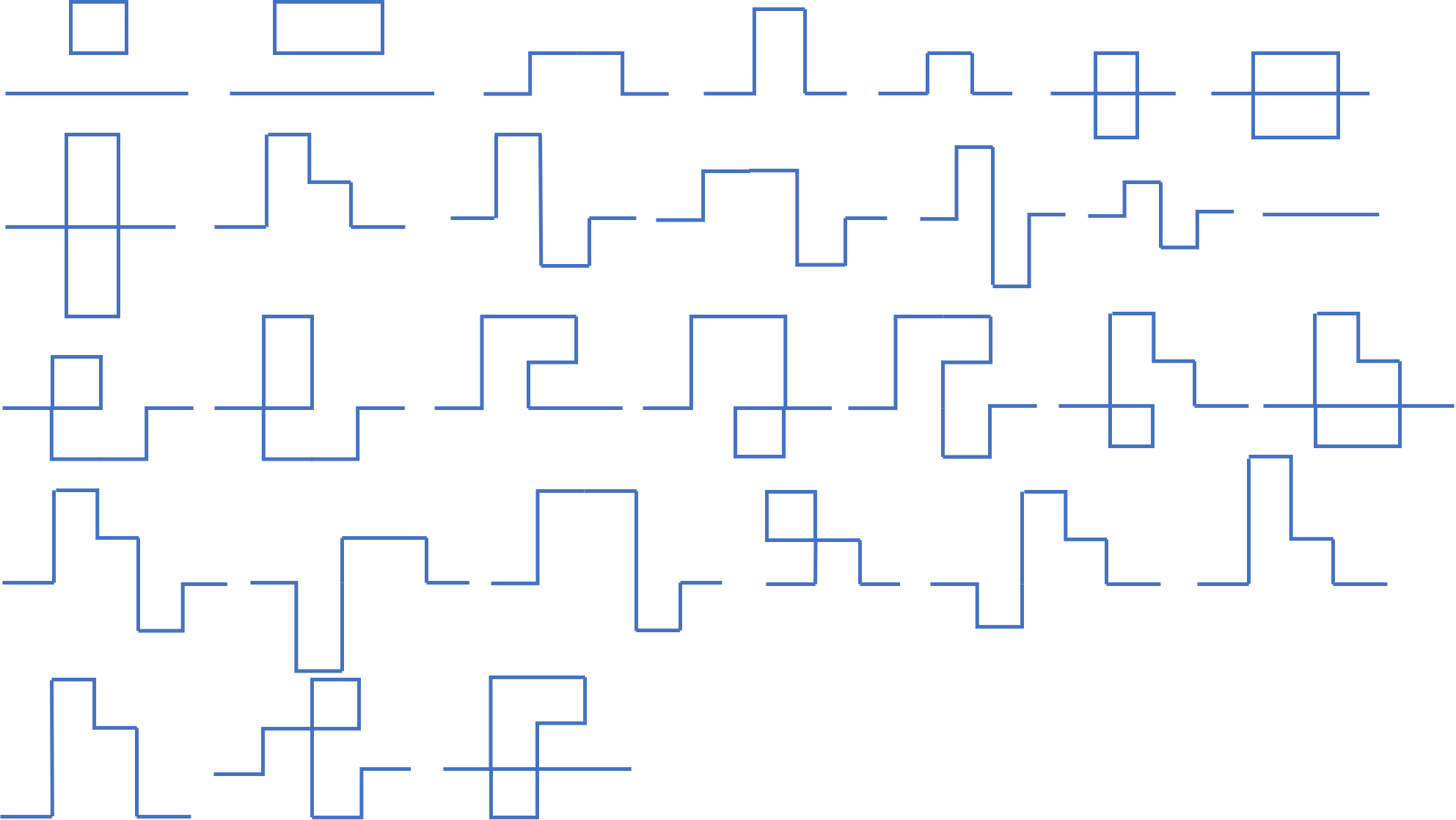}
\centering
\caption{The set of operators used in our simulation.}
\label{fluxtubes}
\end{figure}

Using the blocked links we can now create a basis of Polyakov loop operators of different shapes. In Fig.~\ref{fluxtubes} we present the shapes used in our simulation. We try to use the operators that are as simple as possible, meanwhile including enough operators that break those two parities, especially transverse parity $P_t$, so as to have decent overlaps onto all the sectors we are interested in. Note that some of these operators look like creating a flux tube and an additional glueball rather than just a flux tube excitation. Equivalently, using the $SU(N)$ language, they look like multi trace operators.
  However, for the ${\mathbb Z}_2$ theory there is no sharp distinction between single trace and multi trace operators, because any operator can be formally presented in the single trace form by connecting different components by 
  going back and forward along some path between them (see Fig.~\ref{multitrace}), given the Abelian nature.

Finally, to obtain operators with a definite set of quantum numbers one performs averaging over the action of the corresponding symmetry transformation.  
For example, in order  to construct 
an operator with a definite longitudinal momentum $p$, one sums over all longitudinal translations with a phase
\begin{equation}
    \phi(p) = \sum_{k=1}^L \phi(x+ak) e^{ipak} \,.
\end{equation}
In the same way one constrains $p_{t} = 0$ by summing over all the translations in the transverse $y$ direction without a phase. 

Similarly, one may obtain operators with definite value of transverse and longitudinal parities $(P_t,P_l)$.
For example, as we discussed, at $p=0$ both parities can be be assigned, so we get four different sectors $(++), (+-), (-+)$ and $(--)$.
To construct the corresponding operators one starts with a Wilson line operator $U_C$ corresponding to a certain path $C$, and defines the following eigenstate combination
\be
\tilde{U}_C=(U_C \pm U_{P_lC}) \pm (U_{P_tC} \pm U_{P_tP_lC})\;.
\ee
Here the signs inside the brackets correspond to the eigenvalue of $P_l$, and the sign in the middle corresponds to the eigenvalue of $P_t$.

\section{Results}
\label{section_results}
Let us now present results of our simulations.
In this work we performed ${\mathbb Z}_2$ lattice gauge theory simulations at $\beta = 0.756321$.
This value corresponds to the rough and confining phase. It is sufficiently close to the critical value $\beta_c = 0.7614133(22)$ \cite{blote1999cluster}, to allow for sufficiently long and clear plateaux in the effective mass.
Namely, as follows from the results presented later, for this value of $\beta$ the correlation length $\xi$ (which is set by the inverse mass of the lightest glueball $\xi=m_G^{-1}$) is equal to
\[
\xi=4.631(8) a\;.
\]
Unless specified otherwise, the results presented are obtained on lattices of a size
\[
l_{\perp} = l_t = 70a\;,
\] 
in the transverse and time directions, and the lattice size along the string is varied in the range
\[
R\in [20a,80a]\;,
\]
which corresponds to the range
\[
R\in[1.38\ell_s,5.53\ell_s] \,,
\]
in string units, where the string length is obtained by fitting the absolute ground state energy of the flux tube to the GGRT formula. Recall that the finite temperature deconfinement transition corresponds to $R\sim 0.82 \ell_s$. 
In order to estimate finite volume corrections and for some other checks we also used lattices with other transverse sizes in the range
 from $55a$ to $300a$. 
 These values of  lattice parameters and the corresponding basic physical observables are summarized in Table~\ref{table_parameters}.
 \begin{table}[h]
\begin{center}
  \begin{tabular}{c|c|c|c|c|c}
    \hline \hline
    $\beta$ & $\beta_c$ & $R/a $ & $R_c/a$ &  $a/\ell_s$  & $am_G$ \\  \hline \hline
 $0.756321$    & 0.7614133(22)  & [20,80] &  $\sim11.8$ & 0.0691(1) &  0.2159(4)  \\ \hline
 \end{tabular}
  \caption{Basic parameters of our simulation: the value of the coupling and its critical value, the range of the string circumference and its critical value, the string tension and the lightest glueball mass.}
\label{table_parameters}
\end{center}
\end{table}

%
%
%
%
%

Let us now present results of simulations with these parameters. We start with the absolute flux tube ground state, and continue to excited states in different sectors. 
Comparing the result to the GGRT spectrum we find that the most pronounced qualitative difference is the presence of an extra state in the parity $(++)$ sector at $q=0$. This state can naturally be interpreted as a massive scalar resonance on the string worldsheet. 
We identify the corresponding state also in the $q=1$ sector. Later we present results of an additional dedicated analysis which indicates that this resonance is actually caused by the bulk glueball rather than by a genuine worldsheet state.

\subsection{The absolute ground state and the string tension}
\label{ground_state}
\begin{figure}[t!]
\scalebox{1}{\includegraphics[width=0.69\textwidth]{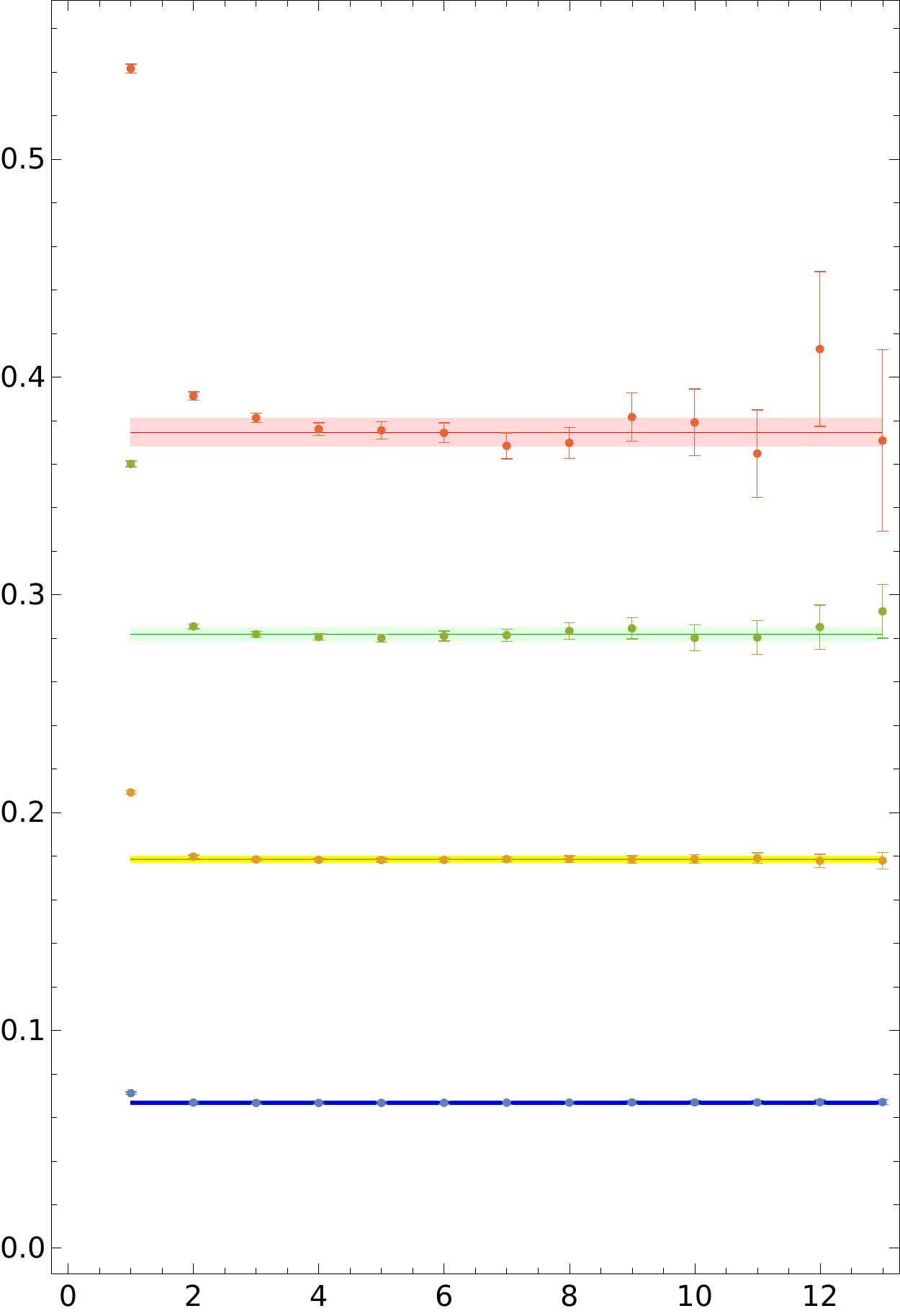}\put(-370,250){ \small $aE(t)$}\put(-170,-20){\small $t/a$}}
\centering
\caption{The effective mass computed as in formula \eqref{effective_mass} as a function of time for the absolute ground states at string circumference $R/a = 20, 40, 60, 80$, represented as blue, yellow, green and red dots. 
The horizontal solid lines  are the resulting fitted values of the state's energies. The shaded bands represent the corresponding $1\sigma$ uncertainty intervals. }
\label{plotemass1}
\end{figure}

The flux tube ground state is translationally invariant, has $q=0$, and belongs to the $(++)$ sector. Understandably, of all the string states this one is the most straightforward to identify.  
As illustrated in Fig.~\ref{plotemass1}, the corresponding effective mass exhibits a well pronounced plateau even for the longest string circumference $R=80a$ considered in our simulations, which allows for a high precision determination of the ground state energy as a function of $R$.

\begin{figure}[t!]
\scalebox{1}{\includegraphics[width=0.5\textwidth]{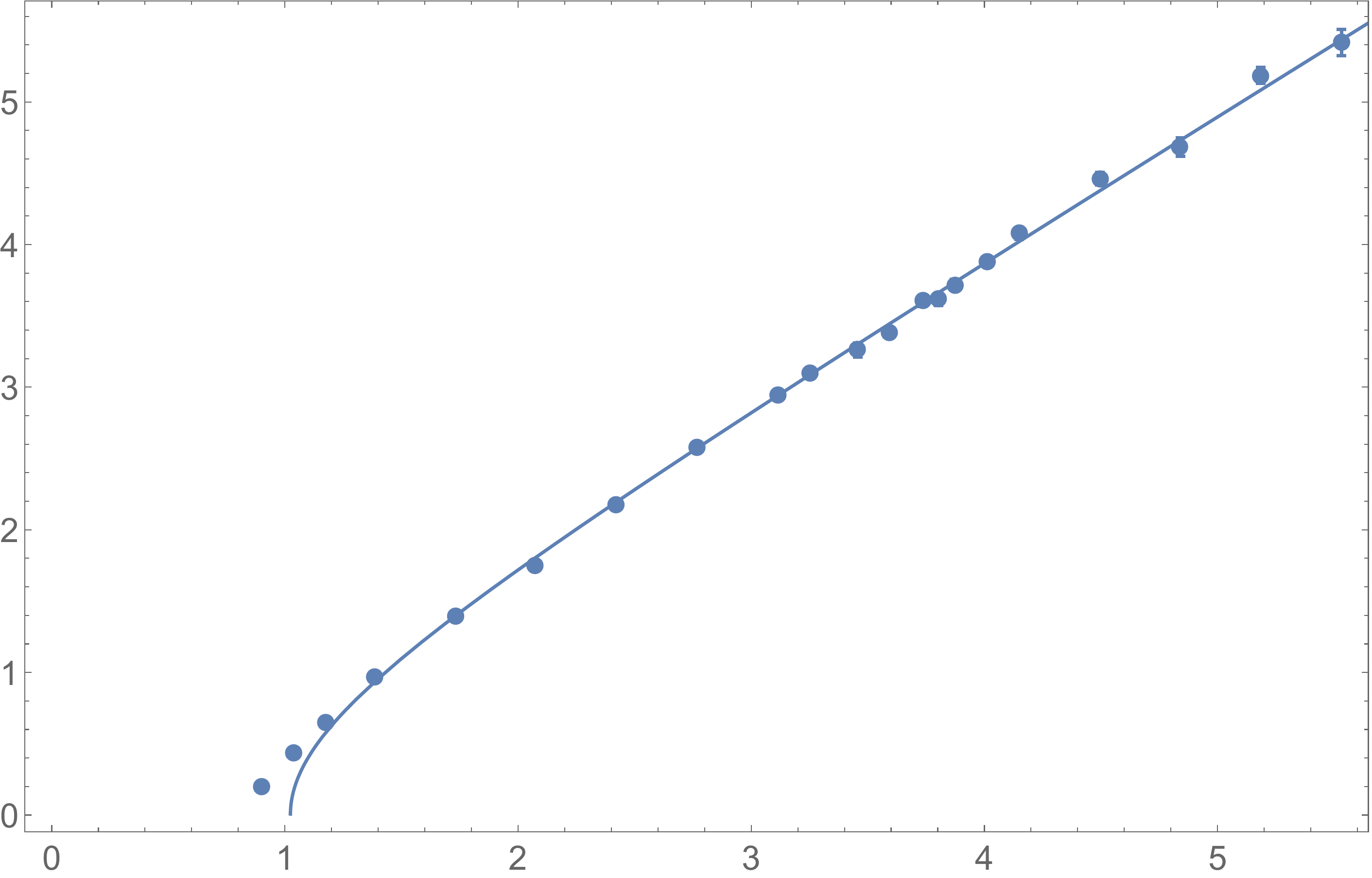}\put(-265,75){ \small $E\ell_s$}\put(-130,-10){\small $R/\ell_s$}}
\centering
\caption{The absolute ground state energy at different string lengths in string units. The solid line is the GGRT approximation for the ground state energy. }
\label{plotgroundstate}
\end{figure}

 In Fig.~\ref{plotgroundstate} we present the ground state energy as a function of circumference $R$.
 The solid line shows the GGRT ground state energy. These results are plotted in string units with the string length parameter $\ell_s$ determined by fitting the data to the GGRT ground state energy.  For the $\ell_s$ extraction we used the data in the range $R \in [25a,80a]$, where the quality of the GGRT fit is the best. The resulting value of $\ell_s$ in lattice units is presented in Table~\ref{table_parameters}. 
  We observe that the GGRT approximation reproduces very well the ground state energy of the Ising string all the way down to $R\sim 1.4\ell_s$. On the other hand, the measured ground state energy
  significantly deviates from the GGRT formula at shorter values of $R$. In particular,  the GGRT ground state energy vanishes at $R\approx 1.02\ell_s$, while the Ising ground state energy stays positive (and approximately linear)  down to a smaller critical value  given by (\ref{RcIsing}). 
 
 To quantify the agreement of the measured ground state energy with the GGRT approximation,  we also fitted the observed energies at the short string regime using the following ansatz 
\begin{equation}
    E_0(R) = E_{GGRT}(R) + {c_\gamma\over\ell_s}\l{\ell_s\over R}\r^\gamma\,,
\end{equation}
 for different values of $\gamma$ and using the string length $\ell_s$ and the coefficient $c_\gamma$ as the fitting parameters. To interpret the results it is instructive to compare the obtained values of $c_\gamma$ with the corresponding coefficients of the $\ell_s/R$ expansion of the GGRT ground state energy itself,
 \begin{equation}
 \label{E0GGRT}
    E_0(R) =  {R\over \ell^2_s} - \frac{\pi}{6} \frac{1}{R} - \frac{\pi^2}{72} \frac{\ell_s^2}{R^3} - \frac{\pi^3}{432} \frac{\ell_s^4}{R^5} + {\cal O}(\ell_s^6)\;,
\end{equation}
where we listed all the universal terms in the $\ell_s/R$ expansion.
For $\gamma=1$, the best fit value of $c_1$ is obtained by fitting within $[1.8l_s,3.5l_s]$ with $\chi^2=1.16$:
\[
c_1\approx-0.009(18)\ll {\pi\over 6}\approx 0.52\;.
\]
The correction $c_1$ we obtain is negligible compared to the value of the corresponding term in (\ref{E0GGRT}), so we conclude that our results provide a quite precise determination of the first universal term in the $\ell_s/R$ expansion (also known as the L\"uscher term).
On the other hand for $\gamma=3$ we obtain by fitting within $[1.4 l_s, 2.8 l_s]$:
\[
c_3\approx0.074(28)\lesssim {\pi^2
\over 72}\approx 0.14\;,
\]
so that our results are consistent with the $1/R^3$ universal term, but cannot be considered as a high precision test of the universality at this order.


%
%
As an additional crosscheck of our simulation we also determined the mass of the lightest glueball $m_G$. When expressed in string units it reads
\be
\label{glueball}
m_G\approx 3.124(10)\ell_s^{-1} \,,
\ee
which agrees well with earlier measurements (cf. (\ref{xitol})).

It is instructive to take a look at the ground state energy for even shorter strings: $R \lesssim 1.2 \ell_s$, as also shown in Figure \ref{plotgroundstate}. Here one observes a large deviation from the GGRT formula. 
Clearly in this regime the $\ell_s/R$ expansion does not converge, so that it cannot be used to measure the perturbative non-universal corrections to the GGRT formula.
It is worth noting that these data do seem to extrapolate towards the deconfining point and exhibit scaling behavior, which indicates that it is a second order phase transition. According to the Svetitsky-Yaffe conjecture \cite{svetitsky1982critical}, this deconfining transition is described by the 2d Ising universality class, of which the scaling behavior is linear
\begin{equation}
    E_0(R) \stackrel{R \rightarrow R_c}{\propto}\left(R-R_c\right).
\end{equation}
From our measurements, it is plausible to be linear. But to really determine the exponent, we need results of higher precision and more data points. One difficulty around the critical point is that the ground state energy goes to zero, so that a larger lattice is needed to perform its accurate determination. 

\subsection{Glueball States}
As a cross-check for our results, we also calculated the low-lying spectrum of ${\mathbb Z}_2$ glueballs in the $0^+$ sector, which is summarized in Table \ref{table_glueball_mass}. 
 Here we can observe the finite volume corrections for low-lying glueball states. For example, for the lightest glueball, the finite volume correction becomes observable for $R \leq 30a$. One may also wonder whether we can observe the state corresponding to two parallel flux tubes, which also has the same quantum numbers. It has the mass of two ground state flux tubes. We do not observe such a state here, which  indicates that the local operators we use for glueball states have poor overlap on these states. Comparing our results with that in \cite{Agostini:1996xy}, our measurements have higher precision, and they agree well. The largest deviation is found for the second excited state, for which our mass is somewhat lower, but still within a $2\sigma$ interval. 
\begin{table}[ht]
\begin{center}
  \begin{tabular}{|c|c|ccc|}
    \hline \hline
    $(l_y/a) \times (l_t/a)$ & $l_x/a$ & \multicolumn{3}{|c|}{$aE$; $0+$} \\  
    \hline \hline
    \multirow{17}{*}{$70 \times 70$} & 25 & 0.1978(28) & 0.2531(91) & 0.3519(75)* \\
    \cline{2-5}
    & 30 & 0.2075(30) & 0.2992(87) & 0.3726(110)* \\
    \cline{2-5}
    & 35 & 0.2163(16) & 0.3635(51) & 0.4528(117)* \\
    \cline{2-5}
    & 40 & 0.2170(15) & 0.3804(81) & 0.5097(95) \\
    \cline{2-5}
    & 45 & 0.2144(17) & 0.3896(54) & 0.5329(100) \\
    \cline{2-5}
    & 47 & 0.2118(14) & 0.3865(96) & 0.5319(115) \\
    \cline{2-5}
    & 50 & 0.2159(20) & 0.3742(62) & 0.5019(166) \\
    \cline{2-5}
    & 52 & 0.2131(22) & 0.3920(52) & 0.5237(93) \\
    \cline{2-5}
    & 54 & 0.2182(12) & 0.3899(89) & 0.5141(93) \\
    \cline{2-5}
    & 55 & 0.2141(20) & 0.3990(40) & 0.5326(108) \\
    \cline{2-5}
    & 56 & 0.2169(18) & 0.3953(44) & 0.5462(59) \\
    \cline{2-5}
    & 58 & 0.2152(22) & 0.3849(66) & 0.4947(158) \\
    \cline{2-5}
    & 60 & 0.2178(20) & 0.3998(52) & 0.5080(154) \\
    \cline{2-5}
    & 65 & 0.2138(20) & 0.3906(64) & 0.5153(182) \\
    \cline{2-5}
    & 70 & 0.2168(11) & 0.3984(61) & 0.5497(67) \\
    \cline{2-5}
    & 75 & 0.2159(17) & 0.4025(44) & 0.5541(66) \\
    \cline{2-5}
    & 80 & 0.2175(17) & 0.3886(73) & 0.5216(144) \\
    \hline
    \multicolumn{2}{|c|}{Fitted masses} & 0.2159(4) & 0.3937(16) & 0.5359(27) \\
    \hline
 \end{tabular}
  \caption{The spectrum of ${\mathbb Z}_2$ glueballs in the $0+$ sector at $\beta = 0.756321$ for different lattice sizes. }  
\label{table_glueball_mass}
\end{center}
\end{table}

\subsection{Excited states}
\label{excited_states}
\begin{figure}[t!]
\scalebox{1}{\includegraphics[width=0.85\textwidth]{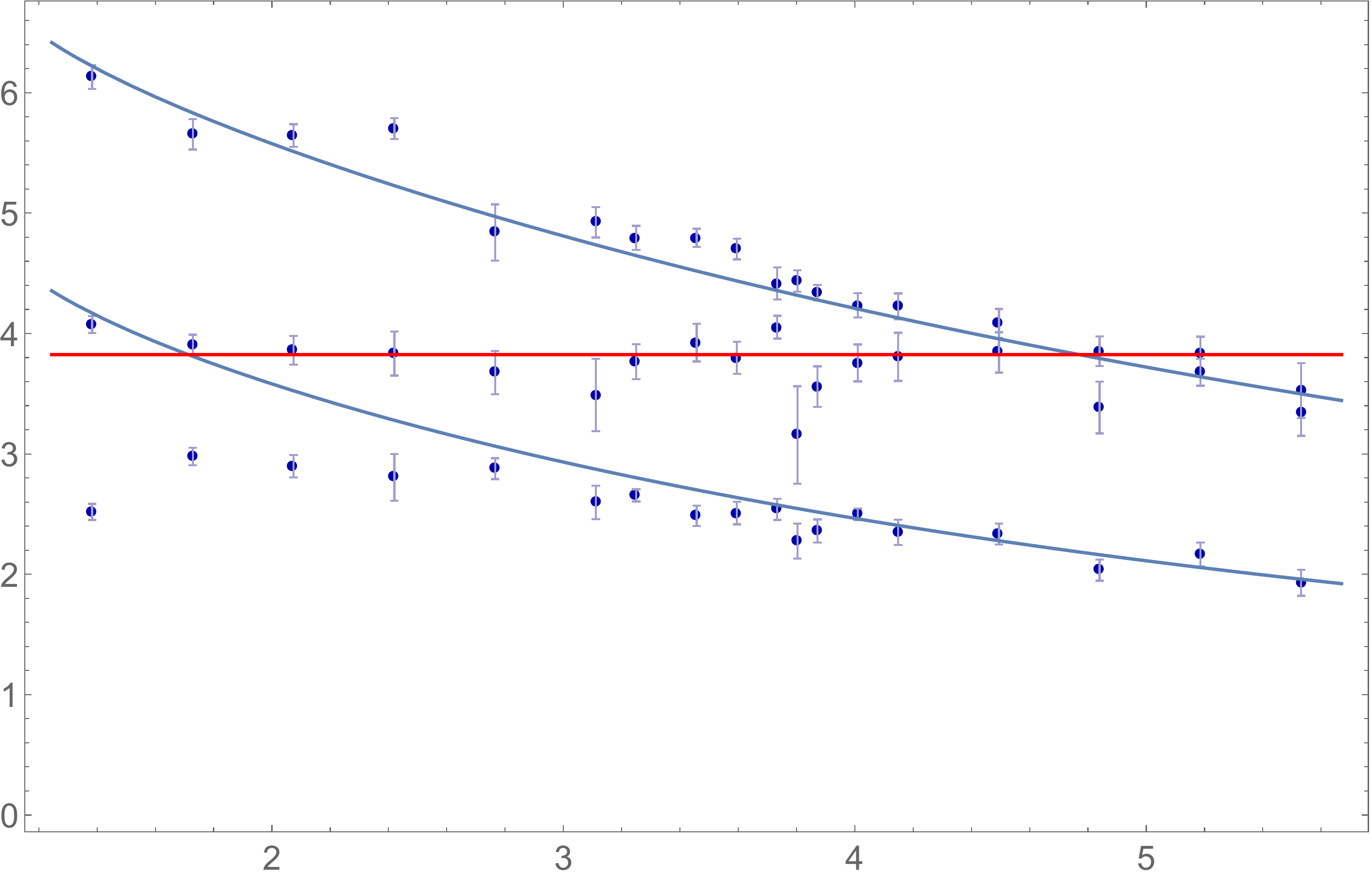}\put(-440,130){ \small $\Delta E\ell_s$}\put(-210,-10){\small $R/\ell_s$}}
\centering
\caption{Energy differences with the ground state for $q=0$ excited states in the $(++)$ parity sector as a function of string circumference  at different string lengths.  Blue curves are the $(1,1)$ and $(2,2)$ GGRT levels. The red horizontal line is the fitted resonance mass. }
\label{plotparityplus}
\end{figure}

Let us now present our results for the excited state's energies of the Ising string. We start with zero momentum states, $q=0$. As discussed before, these states split into four subsectors with different transverse and longitudinal parities,
\be
(P_t,P_l)=(++),(+-),(-+),(--)\;.
\label{parity}
\ee
In Fig.~\ref{plotparityplus} we presented the energy differences between the first three excited states in the $(++)$ sector and the ground state energy. 
As we will see later, restricting to these three states somewhat oversimplifies the overall picture. Nevertheless, it provides a good strating point for interpreting our results.
The numerical values of the corresponding energies (and also of higher excited states) can be found in Table~\ref{table_ppenergy} in the Appendix.
%
In addition to two levels, which are naturally associated with the $(1,1)$ and $(2,2)$ GGRT states\footnote{In the following, for convenience we denote the GGRT levels of states in the format $(N_l,N_r)$. }, we observe on this plot an additional level, which is not associated with any of the GGRT states. Given that the energy gap between this exotic level and the absolute ground state is approximately constant over the large range of $R$, it is natural to associate this state with a massive $(++)$ resonance on a string worldsheet. The resonance mass can be estimated by fitting the energy gap to a constant, which results in
%
\begin{equation}
\label{rgap}
    m\ell_s = 3.825(50)\;,
\end{equation}
where we performed the fit at the intermediate values of string circumference, $R/\ell_s\in [2.4, 4.2]$ to reduce possible effects related to level crossing and  winding corrections.
The latter can be incorporated by applying the TBA technique (cf. \cite{Dubovsky:2014fma}); we will present results of this analysis in a separate publication.
%
%

%

%
%

There are two subtleties worth mentioning here. First, the resonance exhibits two level crossings with the GGRT states in the range of $R$ covered by our data. Namely, it crosses the $(1,1)$ level at $R\sim 2\ell_s$ , and the  $(2,2)$ level at $R\sim 5\ell_s$. In the GGRT spectrum the $(2,2)$ level corresponds to two degenerate states---a two-phonon
and  a four-phonon states. By inspecting Table~\ref{table_ppenergy} one indeed observes two nearly degenerate states close to the  $(2,2)$ level at $R\lesssim 4\ell_s$.
However, one of these states disappears as one approaches the second level crossing at $R\gtrsim 4\ell_s$. The explanation for this is not clear at this point. As follows from  the data presented 
in Table \ref{table_ppenergy}, the energy of the second $(2,2)$ GGRT state starts to increase away from the GGRT spectrum at around $R\gtrsim 3.8\ell_s$. As we will see later the $(++)$ resonance is actually a glueball state mixed with the flux tube. It is possible that these large deviations from the GGRT formula appear above the glueball threshold,  due to interactions between the unbound glueball and the flux tube.
%

The second  subtlety, which is likely related to the first one,  is that the energy gap (\ref{rgap}) is larger than the mass of the lightest glueball (\ref{glueball}) in the infinite volume theory. This implies that  (\ref{rgap}) is not a strictly localized worldsheet state, but rather a metastable bound state between a flux tube and a glueball. In particular, in addition to decaying into a two-phonon flux tube excitation
 it may also decay into a flux tube and a glueball state. Note that the Ising model does not have a parameter which would suppress mixing between genuine flux tube excitations 
 and flux tube states with additional glueballs. This is different from the Yang--Mills case, where such a mixing is suppressed in the 't Hooft large-$N$ limit.  
As a result, one may doubt whether the state (\ref{rgap}) is really due to intrinsic worldsheet dynamics. Perhaps, this state should be considered instead as an admixture of the flux tube and an unbound bulk glueball.
On the other hand, our basis of operators was designed to have a good overlap with states localized in the vicinity of the flux tube, so {\it a priori} one could expect that it is not sensitive to the states with additional unbound glueballs. 
 
We performed several checks to clarify the proper interpretation of this state. First, if the exotic state (\ref{rgap}) were due to an additional unbound glueball, then one would expect to find a state with similar properties also in the $(-+)$ sector.
Indeed, in infinite volume adding a glueball to a flux tube ground state leads to a continuum of states labeled by the asymptotic transverse momentum. In a finite volume this continuum turns into a ``discretuum".
In the absence of interactions between the flux tube and the glueball this discretuum would correspond to the ground state $(++)$ and a series of degenerate doublets with $(++)$ and $(-+)$
parities. However, the interaction with the flux tube breaks the degeneracy, so one obtains a series of alternating $(++)$ and $(-+)$ eigenstates.

Furthermore, energies of all these states, possibly apart  from the lowest  one, have a rather strong dependence on the transverse size $l_\perp$, due to the momentum quantization. This dependence
may be used to distinguish between strongly bound flux tube excitations and unbound states from the discretuum. 

To probe  these states, one may enlarge the set of operators in Fig.~\ref{fluxtubes}
by adding operators which are expected to have a good overlap with unbound flux tube/glueball states, to see whether additional states indeed appear. We will describe the results of this analysis in section~\ref{subsection_probing_glueballs}. As we will see there, our overall conclusion 
is that the state (\ref{rgap}) should indeed be interpreted as a state with an additional unbound glueball.

Let us turn now to excited states in other sectors. For the $q=0$ $(+-)$ sector the effective string theory predicts that the lowest energy state appears at the $(3,3)$ GGRT level and corresponds to a $P_l$ odd linear combination of 
$n_l(3)=1$, $n_r(1)=3$ and $n_r(3)=1$, $n_l(1)=3$ states. Indeed, our analysis does not reveal any low lying states in this sector. We provide the measured energies of the lightest $(+-)$ state in Table~\ref{table_pmenergy} in the Appendix.
At $R/\ell_s\gtrsim 4$ these energies are in between the $(3,3)$ and $(4,4)$ GGRT levels and become significantly heavier at shorter $R$. Given how heavy these states are we expect that their energy determinations are likely to be subject to significant systematic uncertainties. The only robust conclusion one can draw from these results at the moment is that no anomalous light states appear in this sector.

\begin{figure}[t!]
\scalebox{1}{\includegraphics[width=0.85\textwidth]{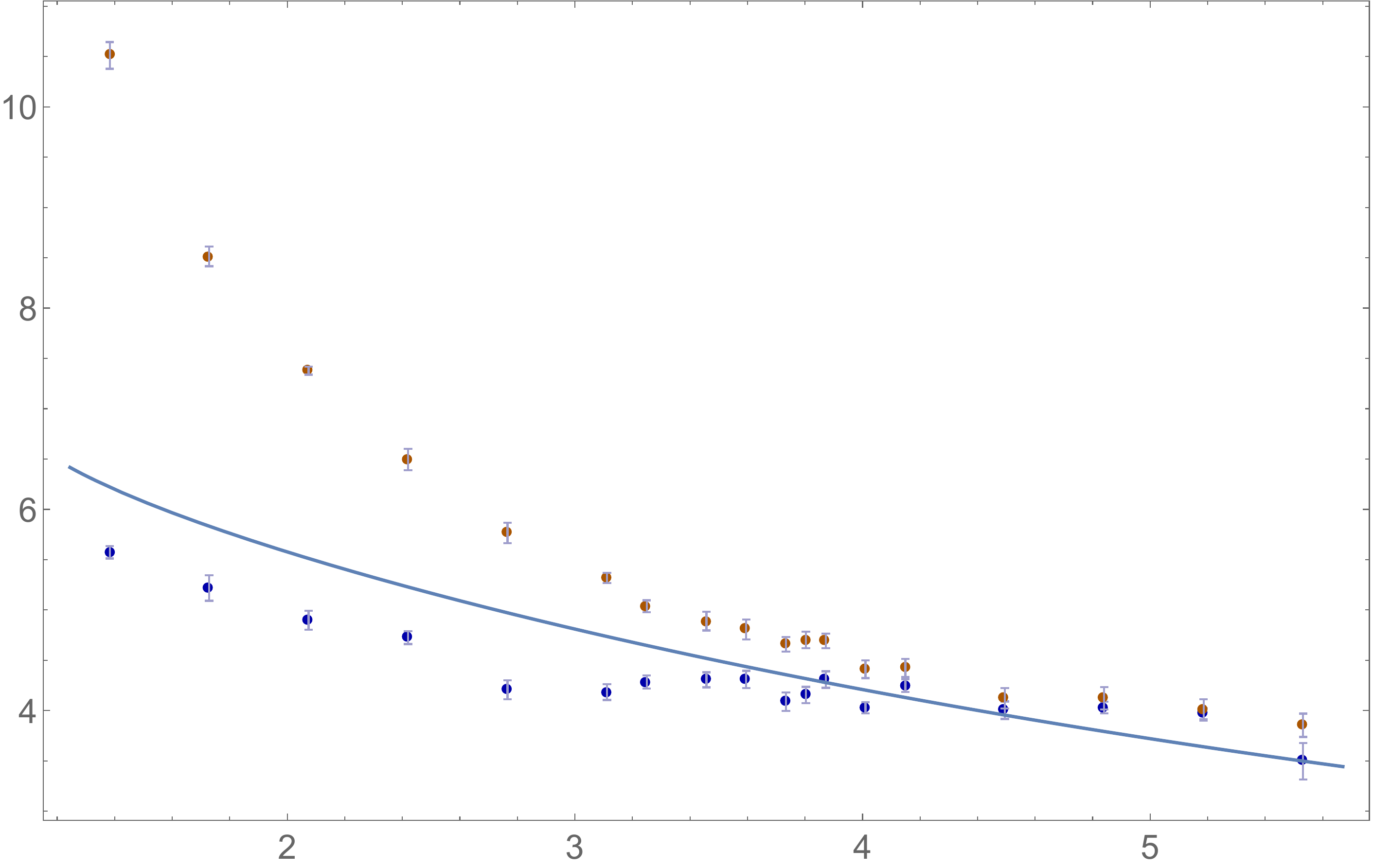}\put(-435,120){ \small $\Delta E\ell_s$}\put(-210,-10){\small $R/\ell_s$}}
\centering
\caption{Energy differences with the ground state for $q=0$ excited states in the $(-+)$ (blue dots) and $(--)$ (brown dots) parity sectors as a function of string circumference  at different string lengths. The blue curve is the energy of the   $(2,2)$ GGRT level.}
\label{plotparityminus}
\end{figure}

Let us discuss now $P_t$ odd states, which are the states with an odd number of phonons. For both $(-+)$ and $(--)$ sectors the lowest GGRT states appear at the $(2,2)$ level, and they correspond to even and odd linear combinations of
$n_l(2)=1$, $n_r(1)=2$ and $n_r(2)=1$, $n_l(1)=2$ states. We plot the measured energies of the lightest states in these sectors in Fig.~\ref{plotparityminus}, and present the numerical values of these energies and those of the heavier states in Tables~\ref{table_mpenergy},~\ref{table_mmenergy} in the Appendix. We observe that at $R\gtrsim 4\ell_s$ these two states are nearly degenerate, as expected for the GGRT spectrum. In this range of $R$ their energies 
are quite close to the expected $(2,2)$ GGRT value, with a minor systematic disagreement. It is most likely due to an overestimate of these rather heavy energies due to an admixture of higher excited states. 

At $R\lesssim 4\ell_s$ the two states are split, and this splitting becomes very large at $R\lesssim 3 \ell_s $, mostly due to a rather dramatic increase in the energy of the $(--)$ state. Interestingly, the energy of the lightest $(+-)$ states discussed earlier exhibits a similar feature in the same range of $R$. At the moment it is hard to tell what is the cause of this effect. 
Note that, as discussed in a similar context in \cite{Chen:2018keo} for the $SU(N)$ data from \cite{Athenodorou:2011rx}, the splitting between three-phonon $(-+)$ and $(--)$ cannot be explained by a correction to the two-phonon phase shift. Instead, it is indicative of a strong inelastic multi-phonon scattering. Interestingly, this splitting appears to be much more dramatic in the Ising case as compared to the $SU(N)$ flux tubes.

\begin{figure}[t!]
\scalebox{1}{\includegraphics[width=0.85\textwidth]{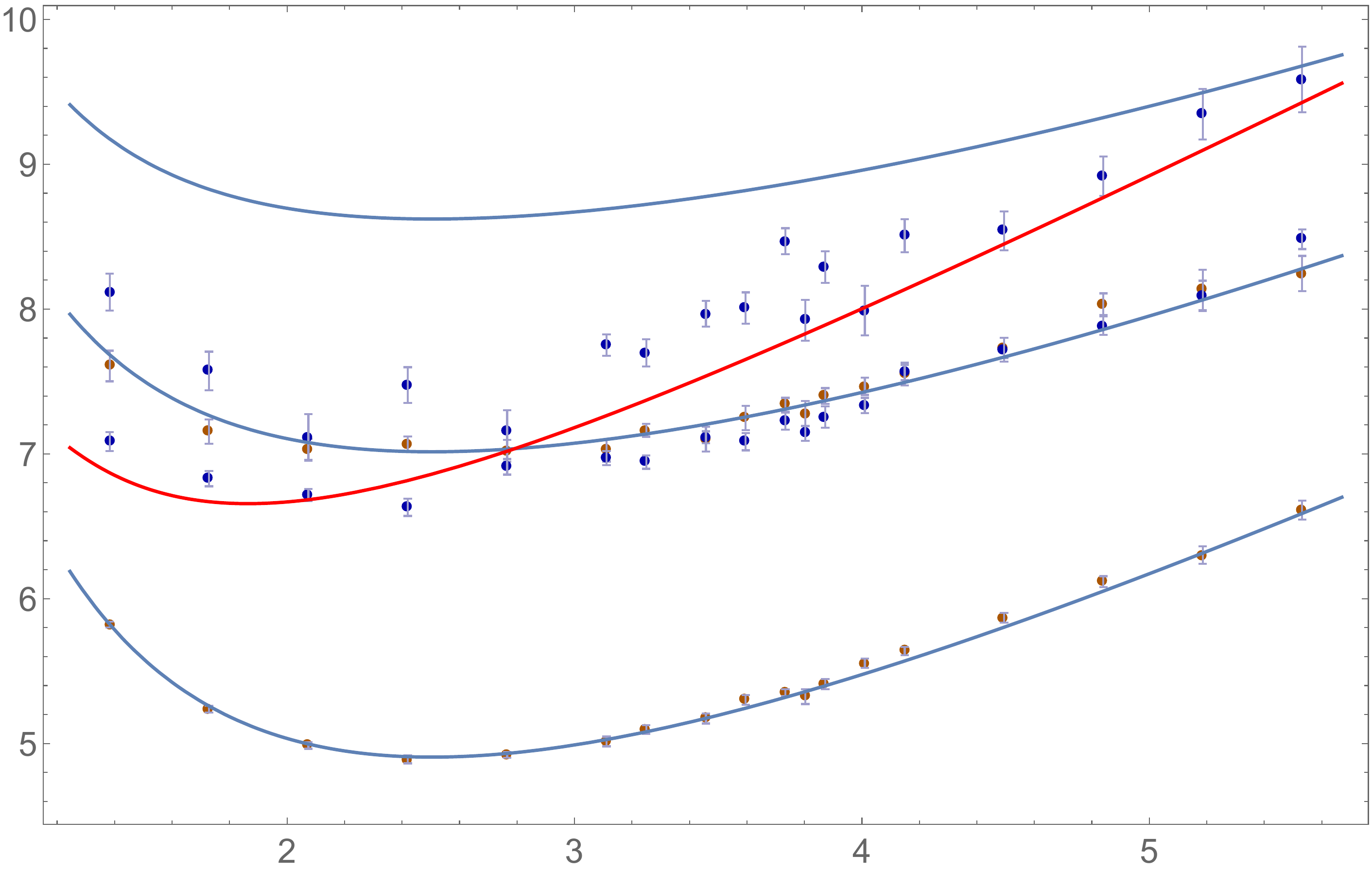}\put(-435,130){ \small $\Delta E\ell_s$}\put(-210,-10){\small $R/\ell_s$}}
\centering
\caption{Energy differences with the ground state for $q=1$ excited states in the $(+)$ (blue dots) and $(-)$ (brown dots) parity sectors as a function of string circumference  at different string lengths. Blue curves show  energies of the   $(1,0)$, $(2,1)$ and $(3,2)$ GGRT levels. A red curve shows an estimate for the resonance state using the resonance mass (\ref{rgap}).}
\label{plotp=1}
\end{figure}

Finally, let us discuss states with nonzero longitudinal momentum $q=1$, which are plotted in Fig.~\ref{plotp=1} and tabulated in Tables~\ref{table_p=1penergy},~\ref{table_p=1menergy}. The ground state in this sector, which is parity odd, agrees exceptionally well with the GGRT $(1,0)$ prediction. This is expected, given that the $(1,0)$ GGRT state corresponds to adding an essentially free (modulo winding corrections) phonon to the ground state of a flux tube.
The first excited parity odd state also agrees very well with the $(2,1)$ GGRT level.

To interpret the two lowest energy parity even $q=1$ states it is instructive to compare their energies to the $(2,1)$ GGRT level and also to the free approximation for the energy of the boosted resonance state,
\begin{equation}
    \Delta E = \sqrt{m^2 + p^2}\,,
\end{equation}
where $p = \frac{2\pi}{R}$. 
We observe from Fig.~\ref{plotp=1} that the two low lying states naturally correspond to a level crossing between the $(2,1)$ GGRT level and a boosted resonance state. 
%

To illustrate how statistical fluctuations influence our results, especially for higher level states, it is instructive to take a look at  the effective mass plateaux behaviour for different states and at the corresponding effective mass fits. In Fig.~\ref{plotemass1} we plotted the effective mass as a function of  time separation for the absolute ground states at different string lengths. As expected, we see that as the string length increases, which corresponds to the heavier ground state energy, statistical fluctuations become larger and the uncertainty in the effective mass determination grows. 
A generic behavior observed for each of the states is that the effective mass exhibits a drop at early times and then stabilizes on a plateau. The rate of the initial drop characterizes the quality of the overlap of our operator basis  onto the corresponding state. Statistical fluctuations increase at larger with time and dominate the measurement at late times. 

\begin{figure}[htbp]
\scalebox{1}{\includegraphics[width=0.68\textwidth]{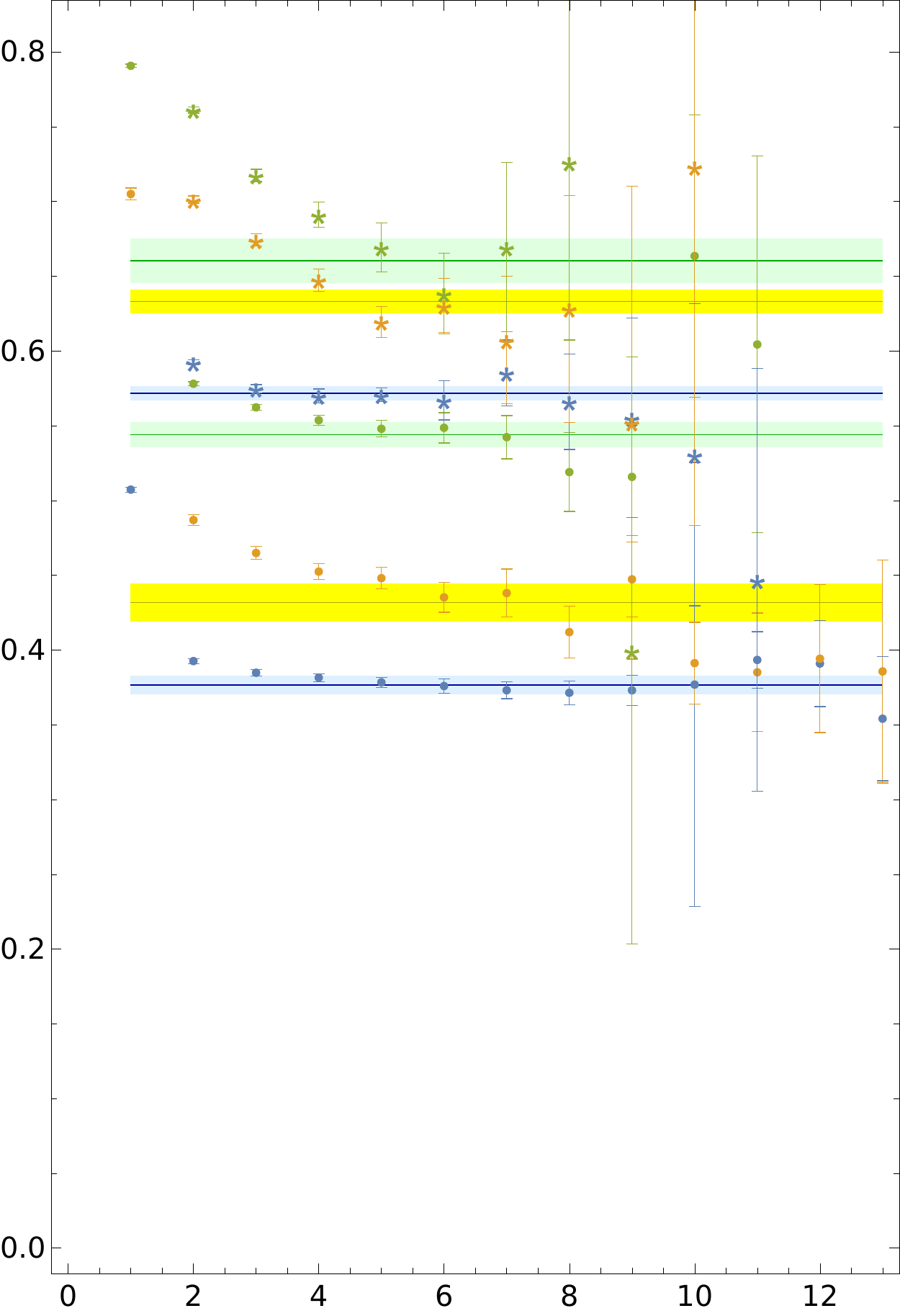}\put(-370,250){\large $aE(t)$}\put(-170,-20){\large $t/a$}}
\centering
\caption{The effective mass computed as in formula \eqref{effective_mass} as a function of time, for the first, second and third excited states in the $q=0$ $(++)$ sector and compactification length $R=40a$, represented as blue, yellow, green dots, and for the ground state, first and second excited states in the  $q=0$ $(-+)$ sector  and compactification length $R=60a$, represented as blue, yellow and green ``$*$''. The horizontal solid lines in dark colors are the fitted value of the mass of the corresponding states. The shaded bands in light colors represents $\pm 1$ standard deviations. }
\label{plotemass2}
\end{figure}

All these features are even more pronounced for excited states as illustrated in Fig.~\ref{plotemass2}. Here we chose the string length such that the non-universal corrections to the GGRT spectrum is small, and at the same time the resonant state is also well pronounced. 
 As compared to the ground state we observe that statistical fluctuations start to dominate the plateau at earlier times. At the energy of around $0.67a^{-1}$, which corresponds to the second excited state in the parity $(-+)$ sector  at  $R=60a$, 
 this effect 
 reaches the point when the position of the plateau is hard to determine. Also, because statistical fluctuations here dominate so early, they are likely to prevent us from observing the point of the plateau stabilization, leading to a possible overestimation of the energy.  Consequently, a reliable spectrum calculation in this energy range requires a larger sample size. 


\subsection{Finite volume corrections}
\label{finite_volume}

Let us discuss the finite size dependence of the presented results. To be more precise, in our simulation we have a finite size lattice system with periodic boundary conditions: $R \times l_{\perp} \times l_t$. The main goal of the simulation is to measure the dependence of string energy levels on the longitudinal size $R$. Instead, in this section we will discuss the sensitivity of the presented results to $ l_{\perp}$ and $l_t$. Our goal is twofold. On the one hand the (in)sensitivity of the measured string energy levels to $ l_{\perp}$ and $l_t$ provide a consistency check for the extrapolation of the measured energy levels to infinite volume. On the other hand, as was already mentioned, the scattering states containing additional glueball(s) states are expected to exhibit a strong dependence on $ l_{\perp}$, which can be used to probe the nature of a massive resonance state observed in the $(++)$ sector.

%
%
%

 In more detail, the spatial finite volume dependence of a single particle or string state with zero momentum in the transverse direction is related to winding corrections associated to (virtual) particles propagating around the spatial circle.
 For massive states, which is always the case for us\footnote{Note that what matters here is the mass of a string as a {\it whole} as it move in the transverse direction. This should not be confused with the mass of longitudinal string {\it excitations}, which is of course zero for the Goldstone modes.}, these corrections are  of order $O(e^{-cam l_{\perp}})$, where the constant $c$ depends on the theory \cite{luscher1986volume}. These corrections are exponentially suppressed, so as we take the transverse size to be moderately large, it will disappear very quickly. 

The story is similar  for corrections associated to the finite size of the temporal circle. To partially account for these corrections we used exponents associated with both directions in time
to fit the two-point correlators instead of a single exponential as written in \eqref{diagonal_correlator}.  This still neglects all the time evolutions that wind around the time circle for more than one round, but these effects are further exponentially suppressed. 

Clearly, winding corrections are  most prominent  for the lightest states.
In particular, $ l_{\perp}$ and $l_t$ need to be sufficiently large for a high precision determination of the low lying string states at small $R$. 
 
For multiparticle scattering states there are larger finite volume corrections that go like ${\cal O}({1}/{(m l_\perp}))$. These are associated with finite momenta of individual particles in a multiparticle state. In particular,  the infinite volume energy spectrum of multiparticle states is continuous. Instead, in a finite spatial volume one expects to find a discretuum  of states which becomes more and more dense as the lattice size increases.


To probe the size of finite volume effects in our results we performed simulations at different lattices and compare the corresponding energy spectra.
We do not find a significant dependence of the measured flux tube spectrum on the temporal lattice size, as follows from the data summarized in Appendix \ref{compilation_spectra}. These data describe  low-lying flux tube spectra measured on $40 \times 55 \times 55$ and $40 \times 55 \times 70$ lattices. The difference is well within error bars. So in what follows we fix $l_t = 70a$, where the time windings can be safely ignored. 

Let us discuss now a set of plots illustrating how energies of low-lying states depend on the transverse size. We do not discuss states in the  $(+-)$  sector  because their energy determinations are not very reliable due to large statistical uncertainties. In this section we fix the size of the longitudinal direction to $R=40a =2.77 l_s$. 
\begin{figure}[ht!]
\scalebox{1}{\includegraphics[width=0.9\textwidth]{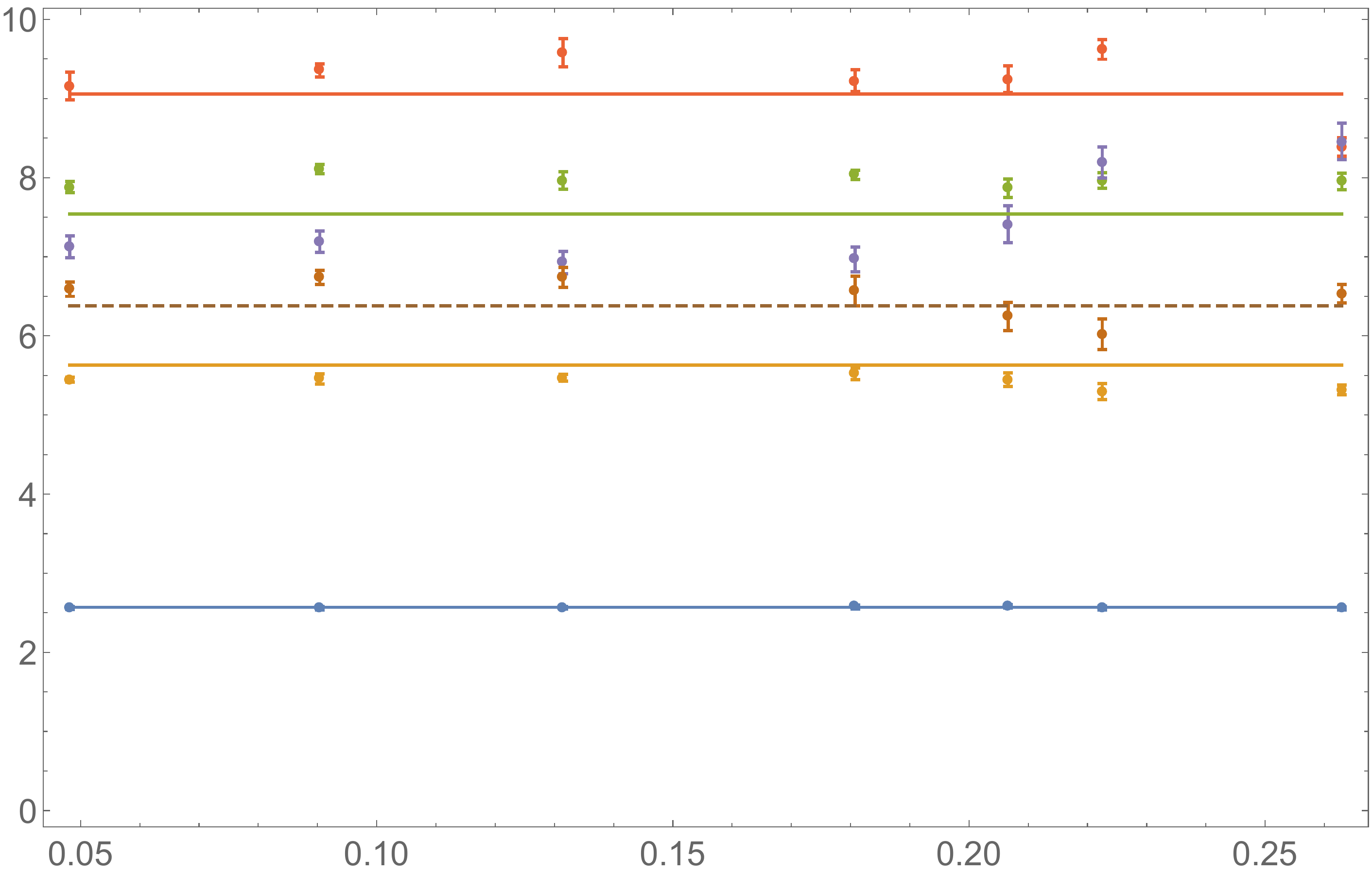}\put(-470,130){ $E/\sqrt{\sigma}$}\put(-230,-20){ $1/l_{\perp}\sqrt{\sigma}$}}
\centering
\caption{Energies in the $q=0$ $(++)$ sector at $R=40a=2.76l_s$ as a function of the inverse transverse size.  Horizontal lines of different colors represent the GGRT spectrum
starting with $N=\tilde{N}=0$. The brown dashed line represents the resonance mass.  }
\label{plot_ppvolume}
\end{figure}

The transverse size dependence of the  $(++)$  states is illustrated in Fig.~\ref{plot_ppvolume}. Blue, yellow, green and red dots are natural to identify with the GGRT states. They match the corresponding GGRT energies fairly well, and do not exhibit strong finite volume dependence. This is also true for the resonance state, which is represented by brown dots. However, there is an extra state represented by purple dots, which exhibits a very pronounced volume dependence at smaller values of $l_{\perp}$.  As follows from our earlier discussion, this volume dependence suggests that this state belongs to a discretuum of scattering states describing a string with an additional glueball with non-vanishing relative momentum.
This suggests that also the resonance state should be zero relative momentum at the bottom of the string-glueball discretuum rather than a genuine string excitation. In the next section we present further evidence supporting this conclusion.

\begin{figure}[t!]
\scalebox{1}{\includegraphics[width=0.9\textwidth]{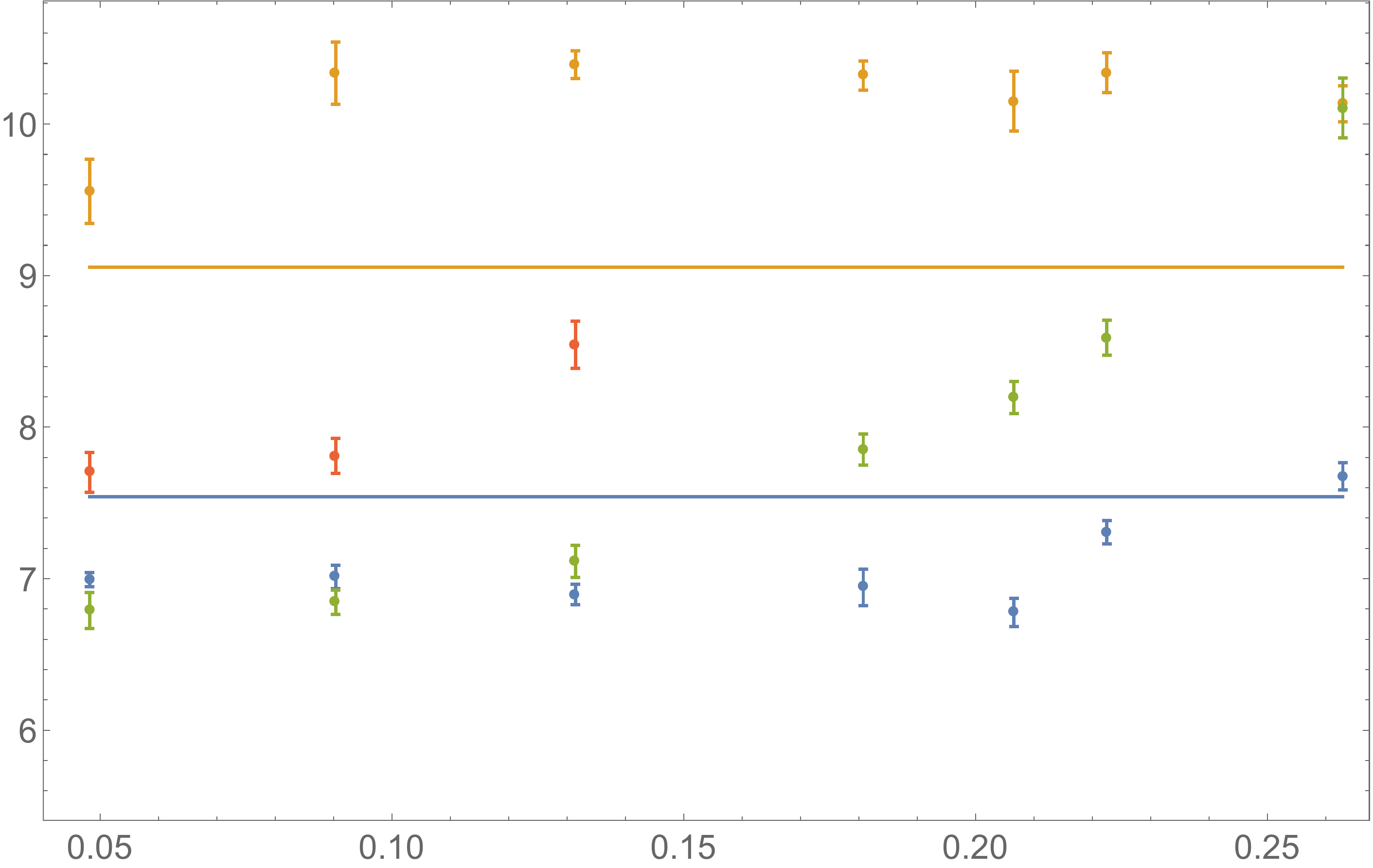}\put(-470,130){ $E/\sqrt{\sigma}$}\put(-230,-20){ $1/l_{\perp}\sqrt{\sigma}$}}
\centering
\caption{Energies in the $q=0$ $(-+)$ sector at  $R=40a=2.76l_s$ as a function of the inverse transverse size. Horizontal lines of different colors represent the GGRT spectrum
starting with $N=\tilde{N}=2$. 
 }
\label{plot_mpvolume}
\end{figure}

The transverse size dependence of the $(-+)$ states is illustrated in Fig.~\ref{plot_mpvolume}. These states are quite a bit heavier than the lightest ones observed in the $(++)$ sector and
it is harder to interpret what happens here. It looks natural to associate  blue and yellow dots with the proper string excitations. Their agreement with the GGRT predictions is not so good, and the lightest (blue) state appears to exhibit some volume dependence at the small values of the transverse size $l_{\perp} \sqrt{\sigma} \lesssim 4.5$. In any case,
one also observes two additional states (green and red) which exhibit a very pronounced volume dependence. As in the $(++)$ case this is suggestive of the scattering states interpretation.


\begin{figure}[htbp]
\scalebox{1}{\includegraphics[width=0.9\textwidth]{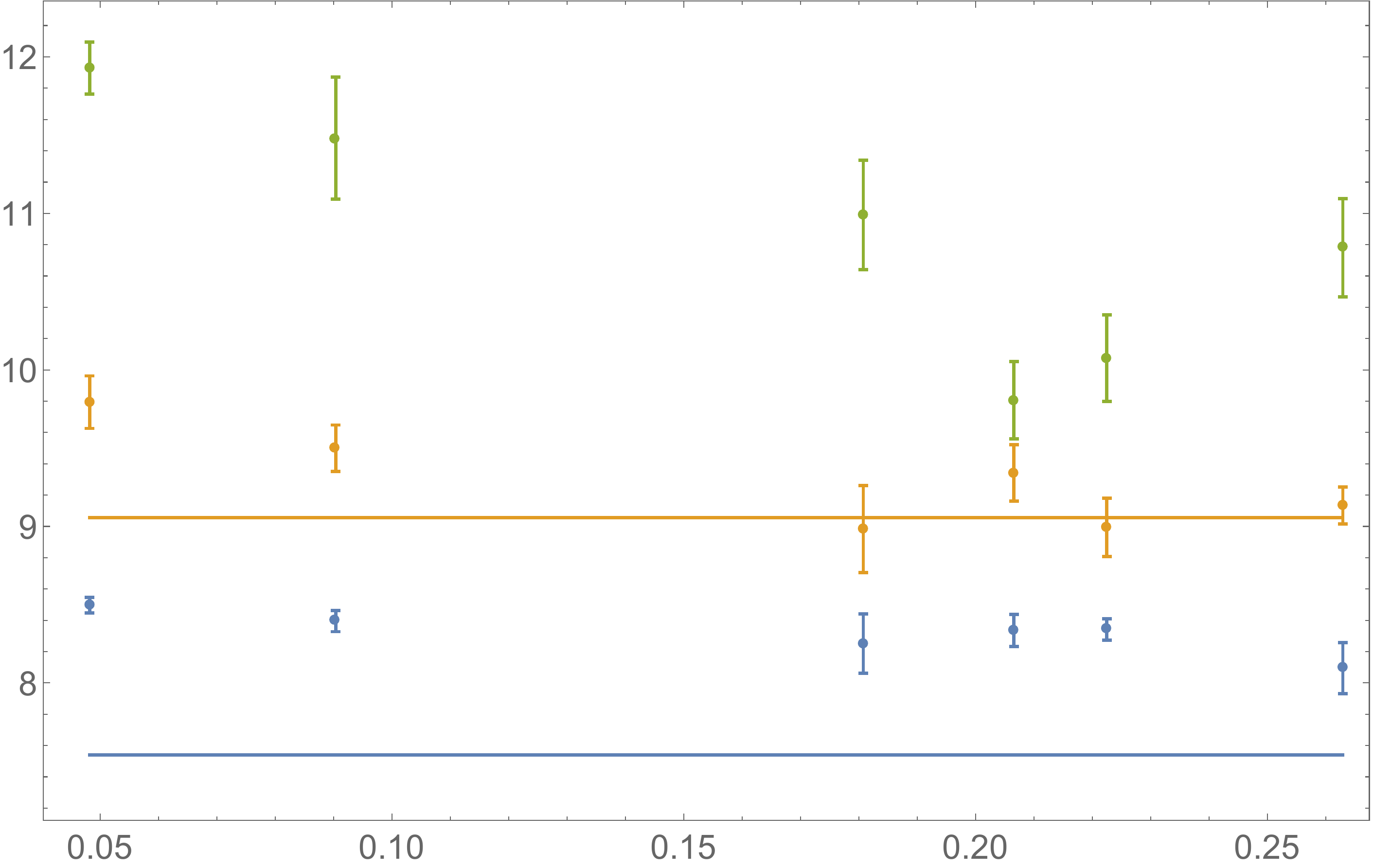}\put(-470,130){ $E/\sqrt{\sigma}$}\put(-230,-20){ $1/l_{\perp}\sqrt{\sigma}$}}
\centering
\caption{Energies in the $q=0$ $(--)$ sector at $R=40a=2.76l_s$ as a function of inverse transverse size.  Horizontal lines of different colors represent the GGRT spectrum
starting with $N=\tilde{N}=2$.}
\label{plot_mmvolume}
\end{figure}

The transverse size dependence of  $(--)$  states  is presented in Fig.~\ref{plot_mmvolume}. There are no recognizable scattering states among the low-lying states
with $E\ell_s\lesssim 10$. This is expected. Indeed to construct a $P_l=-$ scattering states one can either take a  $P_l=-$ flux tube or glueball state, or consider a state where both flux tube and a glueball carry a non-vanishing longitudinal momentum. In all cases the resulting state is expected to be quite heavy.
%

\begin{figure}[htbp]
\scalebox{1}{\includegraphics[width=0.9\textwidth]{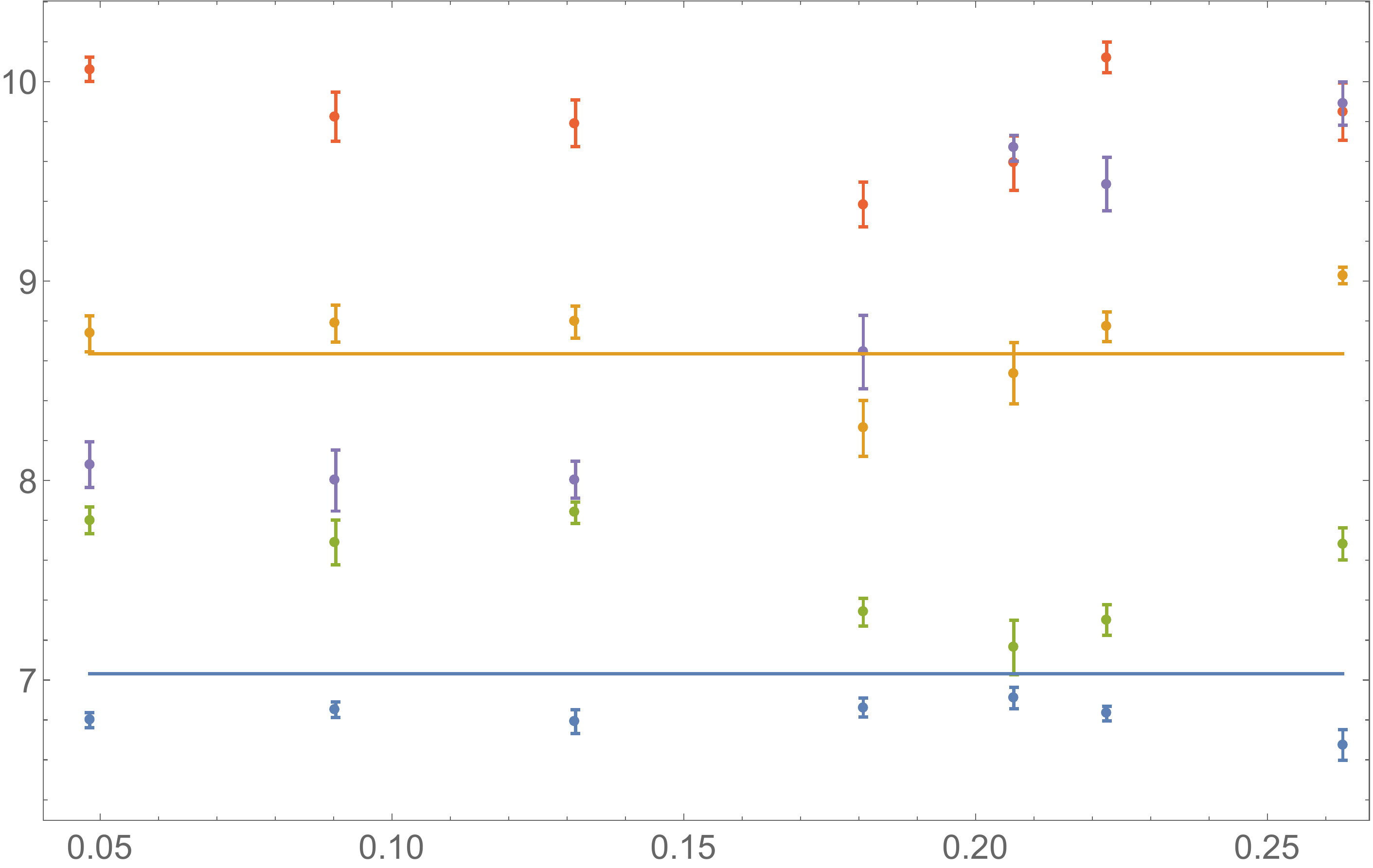}\put(-470,130){ $E/\sqrt{\sigma}$}\put(-230,-20){ $1/l_{\perp}\sqrt{\sigma}$}}
\centering
\caption{Energies in the $q=1$ $(+)$ sector at $R=40a=2.76l_s$ as a function of the inverse transverse size. Horizontal lines of different colors represent the GGRT spectrum
starting from $N=2, \tilde{N}=1$. }
\label{plot_q1pvolume}
\end{figure}

\begin{figure}[htbp]
\scalebox{1}{\includegraphics[width=0.9\textwidth]{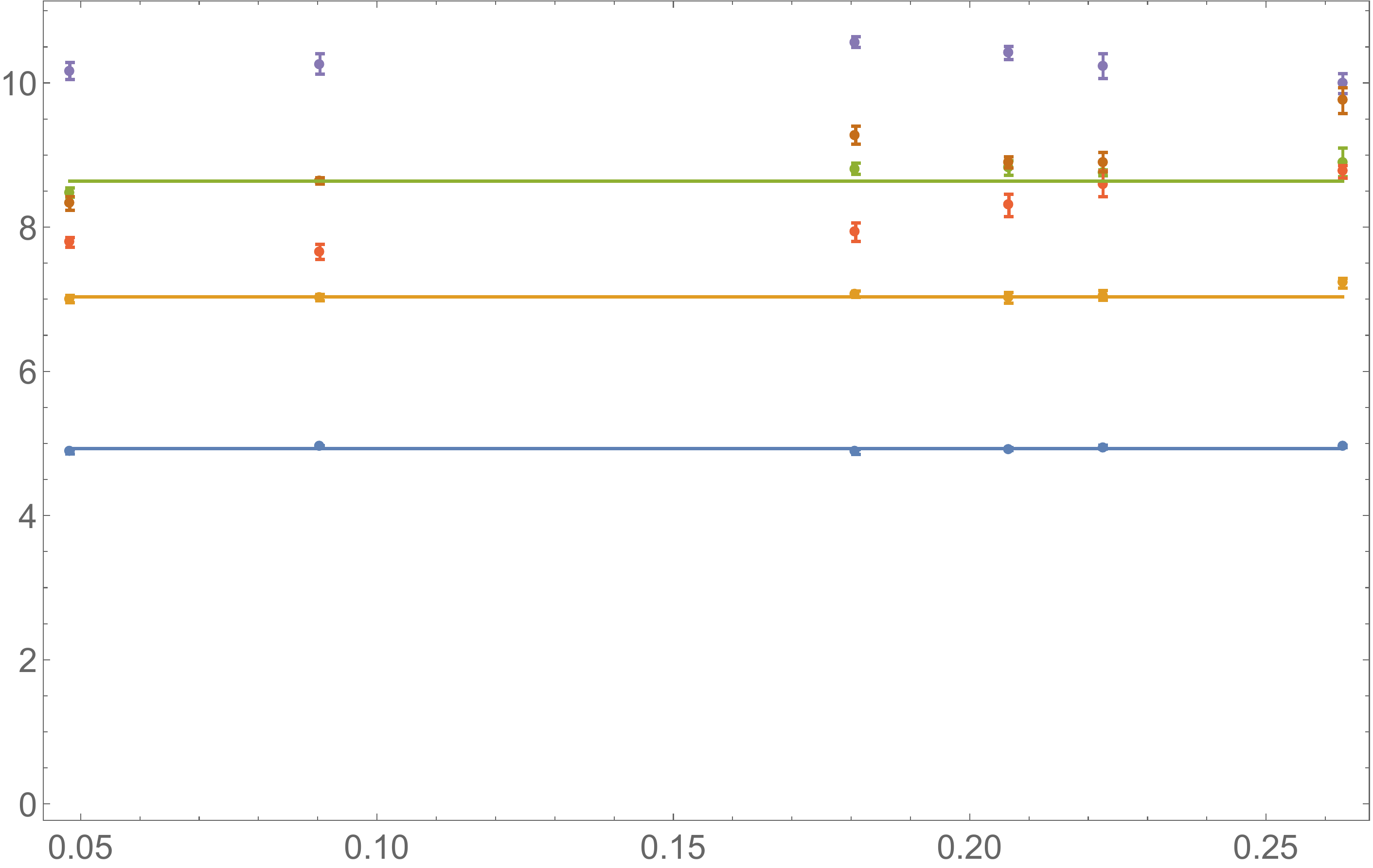}\put(-470,130){ $E/\sqrt{\sigma}$}\put(-230,-20){ $1/l_{\perp}\sqrt{\sigma}$}}
\centering
\caption{Energies in the $q=1$ $(-)$ sector at $ R=40a=2.76l_s$ as a function of the inverse transverse size. Horizontal lines of different colors represent the GGRT spectrum
starting from $N=1, \tilde{N}=0$. }
\label{plot_q1mvolume}
\end{figure}

For completeness we also presented the transverse volume dependence of $q=1$ states in Figs.~\ref{plot_q1pvolume}, \ref{plot_q1mvolume}. The corresponding scattering states can be obtained by boosting a glueball in the $q=0$ 
states, so these states can be used as consistency check. We expect to find scattering states for both $P_t=+$ and $P_t=-$ sectors among $q=1$ states. These states with strong finite volume dependence are indeed present and represented by purple dots in 
Fig.~\ref{plot_q1pvolume} and by red dots in Fig.~\ref{plot_q1mvolume}. The green dots in Fig.~\ref{plot_q1pvolume} represent a resonance state, which can be plausibly reinterpreted as string-glueball discretuum with zero relative momentum.

We conclude that for the coupling $\beta=0.756321$, which we use, a lattice with $l_t= l_{\perp}=70a$ is large enough to ignore finite size effects for the GGRT states at the current level of precision at values of $R$ which is not  too close 
to the deconfining value $R_c = 0.82 \ell_s$.

We do see strong finite volume corrections associated both with $l_t$ and $l_\perp$ dependence as we approach the deconfinement transition $R_c = 0.82 \ell_s$. A much larger lattice size is needed to perform accurate measurements in the vicinity of that point.
Also, we see evidence for the existence of the flux tube-glueball scattering states at large transverse size for both values of the transverse parity $P_t$. This indicates that our set of operators
have a sizable overlap with these states 
 and calls for a more rigorous look on  the nature of the massive state in the $(++)$ sector. This will be the goal of the next section.

\subsection{Including multitrace operators}
\label{subsection_probing_glueballs}
A sizable mixing between flux tube and scattering states is an interesting peculiarity of the Ising model, not present in the non-Abelian Yang--Mills theory.
In the Yang--Mills case the scattering states are created by multitrace operators whose overlap on the flux tube states produced by single trace operators is suppressed even at  moderately large number of colors $N$. As discussed before, in the Ising case there is no distinction between multitrace and single trace operators.
We just saw, this leads to a substantial overlap of our operator basis (which was intended to create pure flux tube states) on the scattering states. On the other hand, this basis is definitely not very well suited for an accurate identification and separation of the scattering states, because one still expects that the corresponding overlap is somewhat suppressed as a consequence of locality. Hence, it should be  instructive to enlarge the operator basis by introducing additional operators with a good overlap on the scattering states. This will allow us to better probe the nature of the $(++)$ resonance and to confirm its interpretation as a zero momentum scattering state. The additional (pseudo) multi trace operators can be constructed by considering a product of (smeared and blocked) plaquette operators $\phi_G$ producing glueball states with the straight Polyakov loop  \eqref{strPol},
\begin{equation}
\label{scat}
    \phi_{scattering} = \sum_{n,m=1}^{l_{\perp}/a} \phi_{P} (y+na) \phi_{G} (y+ma) e^{\frac{2\pi i q_{\perp} (n-m) a}{l_{\perp}}}\;.
\end{equation}
The double sum in (\ref{scat}) is needed to project on a state with a vanishing total transverse momentum, which is also characterized by a relative momentum $q_\perp$\footnote{Note that $q_\perp$ is only an approximate quantum number.}.
We include such operators with $q_{\perp}=0,1,2,3,4$ and $P_t=\pm$ ($q_{\perp} =0$ state only appears in the $P_t=+$ sector). On the other hand, for these operators $P_l=+$  because this holds for the $\phi_G$ and $\phi_P$ that we use, and no relative longitudinal momentum is introduced.

We now repeat the analysis of the transverse volume dependence of the spectrum using this extended basis of operators.
This should allow a more thorough determination of the low-lying spectrum including also the discretuum of scattering states. If the $(++)$ resonance is a genuine string state, one expects to find two low-lying massive states that don't receive pronounced finite volume corrections. One of these states  would then correspond to the lowest lying glueball scattering state and another to the string excitation (which can also be interpreted as a  bound state of a string and a glueball).

\begin{figure}[htbp]
\scalebox{1}{\includegraphics[width=0.9\textwidth]{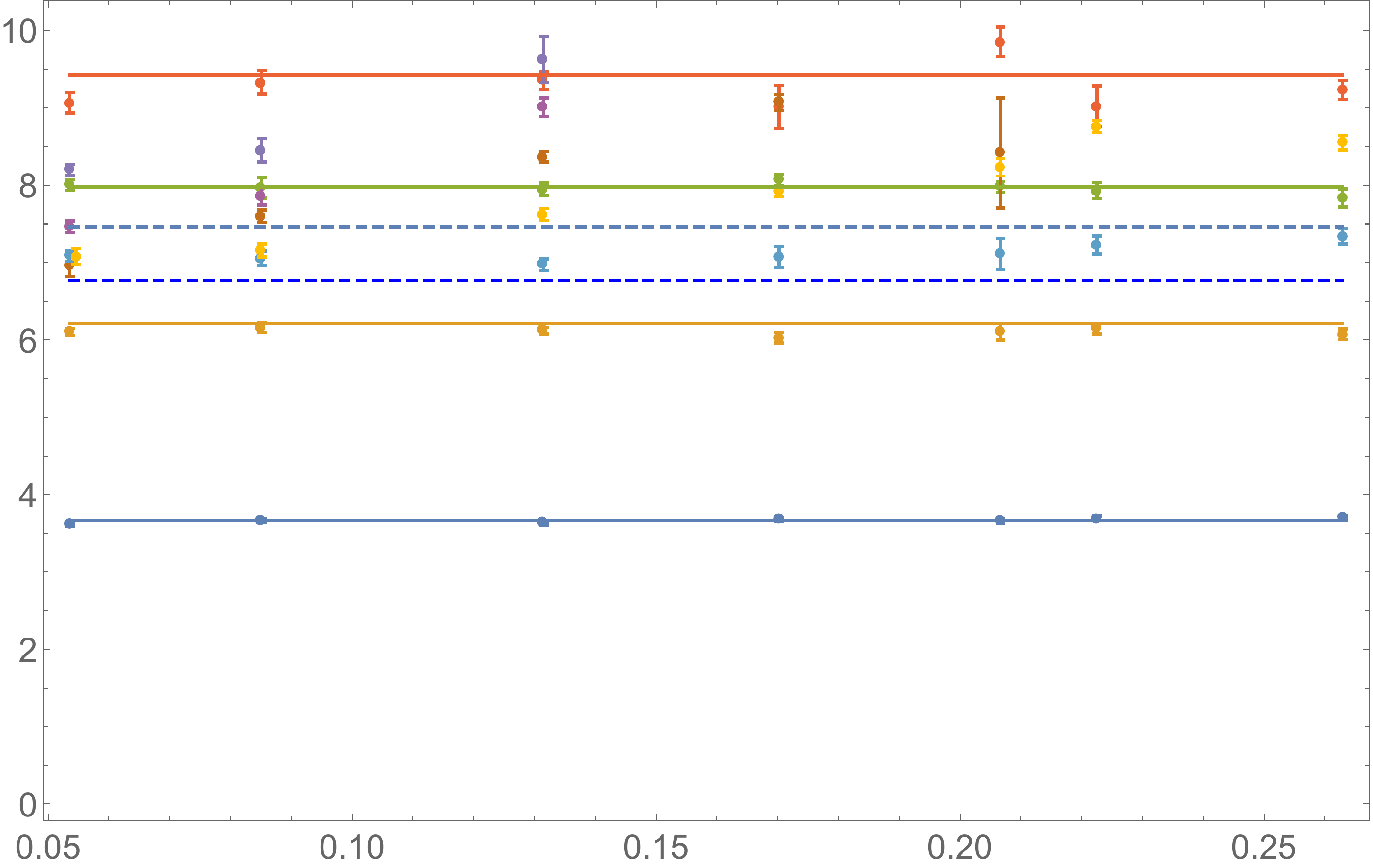}\put(-470,130){ $E/\sqrt{\sigma}$}\put(-230,-20){ $1/l_{\perp}\sqrt{\sigma}$}}
\centering
\caption{Energies in the $q=0$ $(++)$ sector at $R=55a=3.80l_s$ as a function of inverse transverse size
determined using an extended operator basis. Horizontal solid lines of different colors represent the GGRT spectrum
starting from $N=\tilde{N}=0$. 
The lower dashed blue line represents the energy of the absolute ground state plus the glueball mass. The upper dashed blue line represents the absolute ground state plus the resonance mass as given by (\ref{rgap}).}
\label{plot_ppvolume_5glue}
\end{figure}

The results for the  $(++)$ sector are presented in Fig.~\ref{plot_ppvolume_5glue}. Here we chose the compactification radius $R=55a$ to ensure that the lowest scattering state is well separated from the GGRT states. We clearly see that beneath the $(2,2)$ GGRT level, there is only one non-GGRT state (represented by cyan dots) 
whose energy exhibits only a moderate dependence on a transverse size. In addition, there is a series of non-GGRT states with a strong volume dependence (represented by 
 yellow, brown, purple and mauve-blue dots)
which become very dense at large transverse size and accumulate around the expected threshold for the continuum of the scattering states. It is natural to interpret these levels as
fluxtube-glueball scattering states with $q_{\perp}=1,2,3,4$ and the level represented by the cyan dots as a $q_\perp=0$ state at the bottom of the continuum.

Interestingly, this candidate $q_\perp=0$ state still exhibits a noticeable transverse size dependence in the range of $\ell_\perp$ presented in Fig.~\ref{plot_ppvolume_5glue}.
The corresponding energy gap at the shortest values of $\ell_\perp$ is significantly higher than the glueball mass. This is indicative of a considerable repulsive interaction between the glueball and the flux tube.

These interactions appear to be important also for the states with non-zero relative momentum $q$. In particular, {\it a priori} one could have expected that the $\ell_\perp$ dependence of the corresponding energies can be captured by the free dispersion relation,
\begin{equation}
\label{scattering_disp}
    E = \sqrt{m_{flux}^2 + p_{\perp}^2} + \sqrt{m_{glue}^2 + p_{\perp}^2}\;,
\end{equation}
with $p_\perp=2\pi q_\perp/\ell_\perp$.
However, we find that this ansatz does not provide a very good fit, indicative of considerable interactions with the flux tube. These interactions are expected also to affect the GGRT states above the continuum threshold. 
This may actually resolve one of the puzzles encountered earlier. Namely, one expects to find two  states at the $(2,2)$ GGRT level. However, only one such state is present in Fig. \ref{plot_ppvolume_5glue} (the one labeled by green dots). This phenomenon is also observed in Table \ref{table_ppenergy}, where we find out that one of the $(2,2)$ GGRT states start to deviate from GGRT spectrum at $R \gtrsim 3.8 \ell_s$. It appears that a strong mixing between the GGRT and scattering states may provide an explantation for this effect.

\begin{figure}[htbp]
\scalebox{1}{\includegraphics[width=0.9\textwidth]{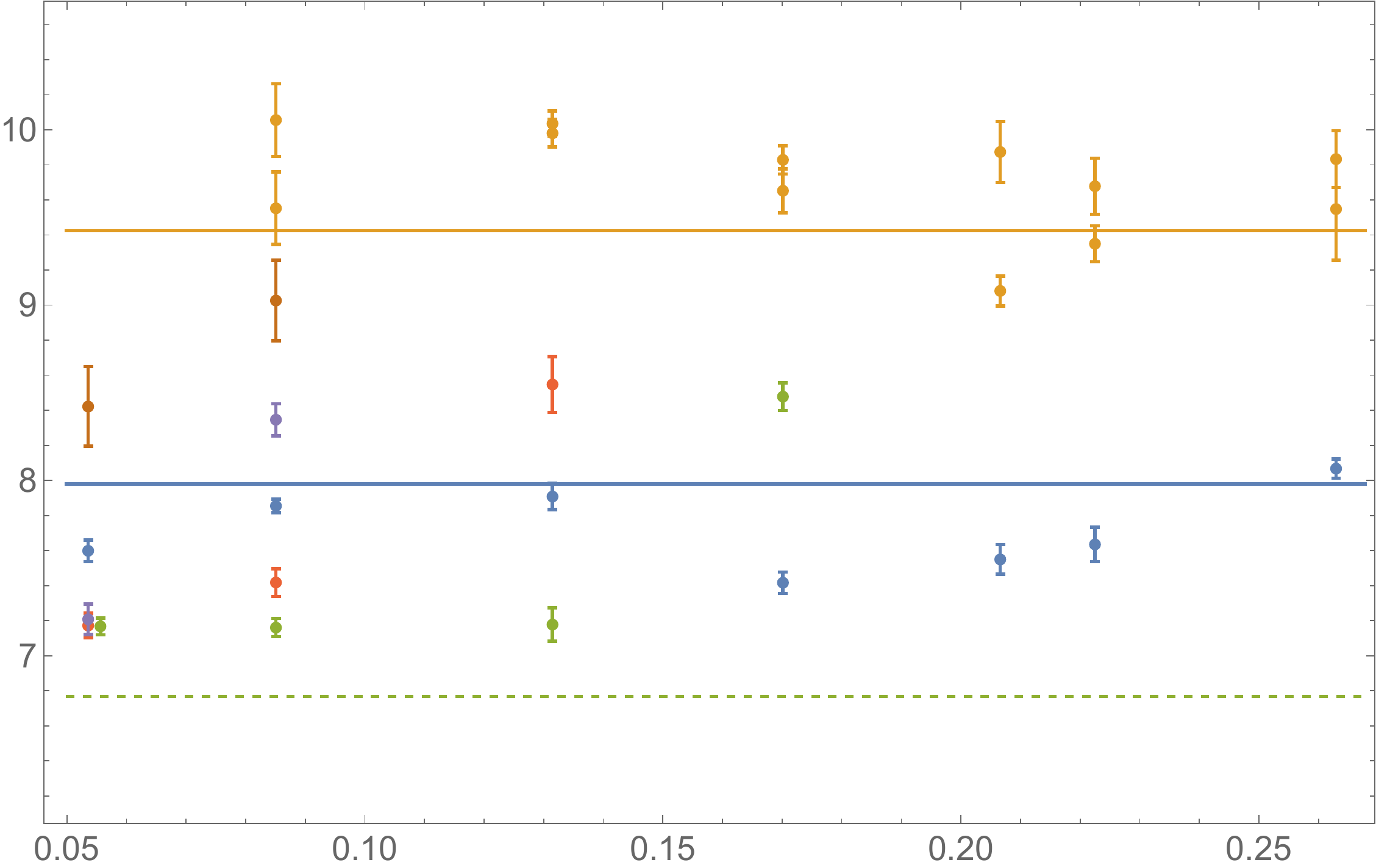}\put(-470,130){ $E/\sqrt{\sigma}$}\put(-230,-20){ $1/l_{\perp}\sqrt{\sigma}$}}
\centering
\caption{
Energies in the $q=0$ $(-+)$ sector at $R=55a=3.80l_s$ as a function of the inverse transverse size
determined using an extended operator basis. Horizontal solid lines of different colors represent the GGRT spectrum
starting from $N=\tilde{N}=2$. 
 The dashed green line represents the energy of absolute ground state plus glueball mass. }.
\label{plot_mpvolume_5glue}
\end{figure}

We observe similar new states with a strong volume dependence also in the $(-+)$ sector.  The corresponding flux tube spectrum as a function of the transverse size is presented in Fig.~\ref{plot_mpvolume_5glue}. Here blue and orange dots are plausible candidates for GGRT $(2,2)$ and $(3,3)$ states given that their volume dependence is relatively mild. In addition we find four states with a strong and monotonic volume dependence, which makes them natural candidates for $q_\perp=1,2,3,4$ states in the discretuum. There is no analogue of the 
$q_{\perp}=0$ state in this sector.  As in the $(++)$ sector the complicated pattern of the corresponding energies, suggests that a considerable mixing between flux  tube and glueball states is present.

%
%

%
%
To summarize, we believe that the analysis presented here strongly disfavors 
the existence of light massive excitations on the worldsheet of the ${\mathbb Z}_2$ confining flux tube. In particular, the state which appears as a massive resonance in the $(++)$ sector corresponds to the glueball scattering state.
In addition, our results indicate the presence of a significant mixing between  flux tube excitations and scattering glueball states. Hence, the effective string theory breaks down above the threshold corresponding to the emission of glueballs.

\section{Concluding Remarks}
\label{section_conclusions}

To summarize, we  calculated the low-lying spectrum of closed flux tube excitations up to the $N=\tilde{N}=3$ GGRT level  in the ${\mathbb Z}_2$ gauge theory, at a coupling $\beta = 0.756321$ which is close to the critical point $\beta_c = 0.7614133(22)$ \cite{blote1999cluster}. The compactification radius covers a wide range $1.38 \ell_s \leq R \leq 5.53 \ell_s$ from moderately short strings to very long ones, but still above the deconfinement transition at $R_c \sim 0.82 \ell_s$ \cite{caselle2003string}. The resulting spectrum agrees with the GGRT predictions for $N=\tilde{N} \leq 1$ states within most of the range of $R$, and also for $N=\tilde{N}=2$ states for moderately long strings.

Somewhat surprisingly, our analysis did not reveal any massive excitations on the worldsheet of the Ising string. A heuristic argument suggesting the presence of a resonance is based on realizing the critical Ising model as an IR fixed point of the $\phi^4$ theory. Then one may attempt to study the properties of the Ising strings by analyzing domain walls in a mass-deformed $\phi^4$ theory. Even though this approach is not based on a well-controlled perturbative expansion at $d=3$, it was argued \cite{munster1990interface} to provide a decent approximation to the ratio of the lighest glueball mass to the string tension. A domain wall in $\phi^4$ theory does support a massive localized resonance \cite{Rajaraman:1982is}, so based on this logic one might have expected to find one also in the Ising case. It will be interesting to study what happens to this resonance using a more systematic approach, based on the $\epsilon$-expansion rather than a direct study of the $d=3$ $\phi^4$ model.

We did observe a state in the $0^{++}$ sector which has an appearance of a massive resonance. However, a detailed analysis revealed that this is a multitrace scattering state with an additional glueball rather than a genuine flux tube excitation. This is related to another interesting (and expected) aspect of the observed spectra. Namely, they indicate the presence of a significant mode mixing between string excitations and glueball scattering states related to the repulsive glueball/string interaction. It will be interesting to perform an analytical analysis of these spectra using an appropriate generalization of the TBA method and to extract the scattering amplitudes describing glueball/string interactions. It will also be interesting to connect this data to the properties of the  line defect in the Ising model at the conformal point, which has been studied in  \cite{billo2013line}.

%

Another possible direction to extend this work is to study the dynamics of strings in the  ${\mathbb Z}_N$ gauge theory. In particular, it will be interesting to study how 
the  3d $U(1)$ gauge theory (studied, e.g.,  in \cite{athenodorou2019spectrum,Caselle:2014eka,Caselle:2016mqu}) is recovered in the $N\to \infty $ limit. It is natural to expect that strong glueball/string interactions should be present in this whole family of theories.

{\it Acknowledgements. } We thank Ofer Aharony, Victor Gorbenko, Michele Caselle, Nabil Iqbal and Yifan Wang for fruitful discussions. This work is supported in part by the NSF grant PHY-2210349, by the BSF grant 2018068 and by the Simons Collaboration on Confinement and QCD Strings. The work of CL is partly supported by funding resources from NYU physics department, and the simulation is run on NYU Greene cluster. 

\begin{appendix}
\section{Compilation of energy spectra}
\label{compilation_spectra}

In this appendix we list all the closed flux tube spectra we've computed for the ${\mathbb Z}_2$ gauge theory at the coupling $\beta = 0.756321$, with different lattice sizes and different quantum numbers. The convention for denoting sectors follows \eqref{parity}. 

\newpage
{\scriptsize
\begin{center}

\begin{longtable}{|c|c|ccccc|}

\hline
$R/a$ & $ l_\perp \times l_t/a^2 $ & \multicolumn{5}{|c|}{ $aE(R) \ ; \ $ $q=0$ $(++)$ } \\ 
\hline
\endfirsthead
\hline
$R/a$ & $l_\perp \times l_t/a^2 $ & \multicolumn{5}{|c|}{ $aE(R) \ ; \ $ $q=0$ $(++)$ } \\ 
\hline
\endhead

\multirow{2}{*}{20}  & \multirow{18}{*}{$70 \times 70$} & 0.0668(8)  & 0.2384(46) & 0.3460(49) & 0.4893(47) & 0.4882(69) \\
& & 0.4996(199)* & 0.6339(109) & & & \\
\cline{1-1}
\cline{3-7}
\multirow{2}{*}{25}  & & 0.0966(11) & 0.3022(50) & 0.3664(58) & 0.4873(87) & 0.5895(290) \\
& & 0.5649(175) & 0.5338(138) & & & \\
\cline{1-1}
\cline{3-7}
\multirow{2}{*}{30}  & & 0.1211(17) & 0.3251(65) & 0.3917(82) & 0.5151(65) & 0.5884(62) \\
& & 0.4929(117) & 0.5838(164)* & & & \\
\cline{1-1}
\cline{3-7}
\multirow{2}{*}{35}  & & 0.1506(13) & 0.3456(134) & 0.4167(126) & 0.5460(60) & 0.5479(135) \\
& & 0.5542(117) & 0.5416(161) & & & \\
\cline{1-1}
\cline{3-7}
\multirow{2}{*}{40}  & & 0.1785(14) & 0.3766(60) & 0.4318(124) & 0.5439(80) & 0.5123(161) \\
& & 0.6392(117) & 0.5854(80)* & & & \\
\cline{1-1}
\cline{3-7}
\multirow{2}{*}{45}  & & 0.2037(19) & 0.3827(97) & 0.4444(208) & 0.5436(87) & 0.5533(242) \\
& & 0.6279(130) & 0.5464(177)* & & & \\
\cline{1-1}
\cline{3-7}
\multirow{2}{*}{47}  & & 0.2143(12) & 0.3969(35) & 0.4737(101) & 0.5448(69) & 0.5596(147) \\
& & 0.6539(64) & 0.6010(69) & & & \\
\cline{1-1}
\cline{3-7}
\multirow{2}{*}{50}  & & 0.2255(34) & 0.4002(58) & 0.4997(108) & 0.5599(52) & 0.5850(154) \\
& & 0.6507(96) & 0.6282(109) & & & \\
\cline{1-1}
\cline{3-7}
\multirow{2}{*}{52}  & & 0.2339(19) & 0.4118(65) & 0.5009(92) & 0.5634(59) & 0.5813(198) \\
& & 0.6407(139) & 0.6327(67)* & & & \\
\cline{1-1}
\cline{3-7}
\multirow{2}{*}{54}  & & 0.2492(27) & 0.4239(61) & 0.5285(66) & 0.5537(92) & 0.6113(122) \\
& & 0.6650(62) & 0.6441(57)* & & & \\
\cline{1-1}
\cline{3-7}
\multirow{2}{*}{55}  & & 0.2500(29) & 0.4106(100) & 0.4715(280) & 0.5600(61) & 0.5612(242) \\
& & 0.6571(100) & 0.6259(72) & & & \\
\cline{1-1}
\cline{3-7}
\multirow{2}{*}{56}  & & 0.2571(26)  & 0.4214(66) & 0.5043(117) & 0.5583(45) & 0.6008(169) \\ 
& & 0.6671(46)* & & & \\
\cline{1-1}
\cline{3-7}
\multirow{2}{*}{58}  & & 0.2686(19) & 0.4409(33) & 0.5278(106) & 0.5609(70) & 0.6136(139) \\
& & 0.6375(113)* & & & & \\
\cline{1-1}
\cline{3-7}
\multirow{2}{*}{60}  & & 0.2819(29) & 0.4404(72) & 0.5412(138) & 0.5701(75) & 0.6386(207) \\
& & 0.6543(88) & 0.7048(96) & & & \\ 
\cline{1-1}
\cline{3-7}
\multirow{2}{*}{65}  & & 0.3085(31) & 0.4640(60) & 0.5683(116) & 0.5850(83) & 0.6663(122) \\ 
& & 0.6567(137)* & 0.6680(199)* & & & \\ 
\cline{1-1}
\cline{3-7}
\multirow{2}{*}{70}  & & 0.3238(45) & 0.4678(61) & 0.5612(149) & 0.5935(85) & 0.6718(202)* \\
& & 0.6483(144)* & 0.6858(174)* & & & \\ 
\cline{1-1}
\cline{3-7}
\multirow{2}{*}{75}  & & 0.3586(38) & 0.5012(68) & 0.6167(96) & 0.6058(77) & 0.7401(114) \\
& & 0.6921(121)* & 0.7561(118)* & & & \\
\cline{1-1}
\cline{3-7}
\multirow{2}{*}{80}  & & 0.3745(64) & 0.5093(75) & 0.6069(132) & 0.6197(157) & 0.7463(152)* \\
& & 0.7849(126)* & & & & \\ 
\hline
\multirow{2}{*}{40}  & \multirow{2}{*}{$55 \times 55$} & 0.1731(19) & 0.3514(57) & 0.4407(90) & 0.5443(86) & 0.5715(177) \\
& & 0.5694(118) & 0.6240(146) & & & \\
\hline
\multirow{2}{*}{40}  & \multirow{2}{*}{$55 \times 70$} & 0.1766(14) & 0.3678(43) & 0.4517(82) & 0.5497(72) & 0.5847(159) \\
& & 0.5800(81) & & & & \\ 
\hline
\multirow{2}{*}{40}  & \multirow{2}{*}{$65 \times 70$} & 0.1772(17) & 0.3662(70) & 0.4162(133) & 0.5506(69) & 0.5665(136) \\
& & 0.6653(87) & 0.5415(176) & & & \\
\hline
\multirow{2}{*}{40}  & \multirow{2}{*}{$80 \times 70$} & 0.1780(17) & 0.3817(50) & 0.4540(130) & 0.5556(40) & 0.4818(108) \\
& & 0.6378(96) & 0.6385(98) & & & \\
\hline
\multirow{2}{*}{40}  & \multirow{2}{*}{$160 \times 70$} & 0.1768(15) & 0.3772(44) & 0.4660(61) & 0.5607(41) & 0.4972(94) \\
& & 0.6469(57) & 0.5515(104)* & & & \\
\hline
\multirow{2}{*}{40}  & \multirow{2}{*}{$300 \times 70$} & 0.1770(13) & 0.3767(21) & 0.4557(63) & 0.5449(48) & 0.4928(96) \\
& & 0.6332(122) & 0.5611(103)* & & & \\
\hline 
\caption{The energies, $E(R)$, of the lightest seven flux tube states with 
length $R$ in the sector $q=0$ $(++)$. } 
\label{table_ppenergy}
\end{longtable}

\end{center} }

\begin{table}[h]
\begin{center}
\begin{tabular}{|c|c|c|}\hline
$R/a$ & $l_\perp\times l_t /a^2 $ &  $aE(R) \ ; \ $ $q=0$ $(+-)$ \\ \hline
20  & \multirow{18}{*}{$70 \times 70$} & 0.9337(259) \\
25  & & 0.8790(60) \\
30  & & 0.8055(221) \\
35  & & 0.7678(83) \\
40  & & 0.7155(172) \\
45  & & 0.7316(80)* \\
47  & & 0.7392(99)*  \\
50  & & 0.7520(115) \\
52  & & 0.7487(105) \\
54  & & 0.6905(208) \\
55  & & 0.7501(123)* \\
56  & & 0.7272(85) \\
58  & & 0.7644(55)* \\
60  & & 0.7189(112)* \\ 
65  & & 0.7264(132)* \\ 
70  & & 0.7528(53)* \\ 
75  & & 0.7401(159)* \\
80  & & 0.7242(126)* \\ 
40  & $55 \times 55$ & 0.7458(198) \\
40  & $55 \times 70$ & 0.7513(183) \\ 
40  & $65 \times 70$ & 0.7518(79) \\
40  & $80 \times 70$ & 0.7478(80) \\
40  & $160 \times 70$ & 0.7215(185) \\
40  & $300 \times 70$ & 0.7389(79) \\
\hline
\end{tabular}
\caption{The energies, $E(R)$, of the lightest flux tube state with 
length $R$ in the sector $q=0$ $(+-)$.}
\label{table_pmenergy}
\end{center}
\end{table}

\begin{table}[h]
\begin{center}
\begin{tabular}{|c|c|ccccc|}\hline
$R/a$ & $l_\perp\times l_t /a^2 $ & \multicolumn{5}{|c|}{ $aE(R) \ ; \ $ $q=0$ $(-+)$ } \\ \hline
20  & \multirow{18}{*}{$70 \times 70$} & 0.4497(43) & 0.6023(116) & 0.6516(268) & 0.9297(72) & \\
25  & & 0.4572(88) & 0.5693(87) & 0.7823(110) & 0.8336(357) & \\
30  & & 0.4634(65) & 0.5525(107) & 0.7294(197) & 0.7506(230)* & \\
35  & & 0.4784(45) & 0.5851(41) & 0.7281(41) & 0.7736(329) & \\
40  & & 0.4686(65) & 0.5666(73) & 0.7019(137) & 0.7166(222)* & \\
45  & & 0.4925(54) & 0.5682(177) & 0.6753(100) & 0.7935(123) & \\
47  & & 0.5095(46) & 0.5961(98) & 0.6698(121) & 0.7487(241)* & \\
50  & & 0.5261(52) & 0.5991(87) & 0.6866(71) & 0.7826(102) & \\
52  & & 0.5364(59) & 0.6139(58)* & 0.6904(88) & 0.7770(115) & \\
54  & & 0.5310(64) & 0.6063(100)* & 0.6897(69) & 0.8138(53) & \\
55  & & 0.5405(56) & 0.6393(63) & 0.6994(91) & 0.7418(258)* & \\
56  & & 0.5561(57) & 0.6405(58) & 0.6923(84) & 0.7685(137)* & \\
58  & & 0.5467(39) & 0.6314(62) & 0.6979(83) & 0.7007(83) & \\
60  & & 0.5717(44) & 0.6331(78) & 0.6604(148)* & 0.8186(61) & \\ 
65  & & 0.5795(60) & 0.6652(71)* & 0.6964(108)* & & \\ 
70  & & 0.6059(40) & 0.7006(330) & 0.7084(104)* & & \\ 
75  & & 0.6259(38) & 0.6874(218)* & 0.7141(98)* & & \\
80  & & 0.6177(125) & 0.7305(118)* & 0.7384(73)* & & \\ 
40  & $55 \times 55$ & 0.5278(42) & 0.6653(115) & 0.7093(101) & & \\
40  & $55 \times 70$ & 0.5308(62) & 0.6988(136) & 0.7007(83) & & \\ 
40  & $65 \times 70$ & 0.5052(53) & 0.5939(80) & 0.7149(90) & 0.7571(208) & \\
40  & $80 \times 70$ & 0.4801(83) & 0.5430(71) & 0.7136(66) & 0.6882(161) & \\
40  & $160 \times 70$ & 0.4848(53) & 0.4733(55) & 0.5400(80)* & 0.7147(142) & 0.5930(135)* \\
40  & $300 \times 70$ & 0.4836(32) & 0.4694(82) & 0.5325(92)* & 0.6608(146) & \\
\hline
\end{tabular}
\caption{The energies, $E(R)$, of the lightest four flux tube states (for $40 \times 160 \times 70$ it is five) with 
length $R$ in the sector $q=0$ $(-+)$.}
\label{table_mpenergy}
\end{center}
\end{table}

\begin{table}[h]
\begin{center}
\begin{tabular}{|c|c|ccc|}\hline
$R/a$ & $l_\perp\times l_t /a^2 $ & \multicolumn{3}{|c|}{ $aE(R) \ ; \ $ $q=0$ $(--)$} \\ \hline
20  & \multirow{18}{*}{$70 \times 70$} & 0.7911(92) & 0.8396(220) & 0.8715(63) \\
25  & & 0.6850(68) & 0.7588(50) & 0.7775(99) \\
30  & & 0.6349(27) & 0.7458(84) & 0.9011(484) \\
35  & & 0.6008(74) & 0.6935(170) &  \\
40  & & 0.5763(71) & 0.6459(125) & 0.6780(171) \\
45  & & 0.5709(35) & 0.6772(161) & 0.7331(108) \\
47  & & 0.5615(41) & 0.6669(127) & 0.7235(166)  \\
50  & & 0.5664(64) & 0.6833(121) & 0.7018(168)* \\
52  & & 0.5707(69) & 0.6595(131) & 0.7488(163) \\
54  & & 0.5704(51) & 0.6867(87) & 0.7153(130)* \\
55  & & 0.5784(56) & 0.6894(150) & 0.7056(159)* \\
56  & & 0.5827(51) & 0.6697(186) & 0.7709(76)* \\
58  & & 0.5731(61) & 0.7214(104) & 0.7735(71)* \\
60  & & 0.5838(62) & 0.6984(169) & 0.8113(54)* \\ 
65  & & 0.5877(70) & 0.7686(62) & 0.7783(365)* \\ 
70  & & 0.6121(77) & 0.7115(210)* & \\ 
75  & & 0.6286(72) & 0.6914(222)* &  \\
80  & & 0.6424(80) & 0.7357(326)* & \\ 
40  & $55 \times 55$ & 0.5763(50) & 0.6136(100) & 0.7731(133) \\
40  & $55 \times 70$ & 0.5597(112) & 0.6315(82) & 0.7454(217) \\ 
40  & $65 \times 70$ & 0.5768(47) & 0.6218(129) & 0.6967(191) \\
40  & $80 \times 70$ & 0.5706(131) & 0.6211(192) & 0.7599(241) \\
40  & $160 \times 70$ & 0.5805(47) & 0.6568(103) & 0.7938(270) \\
40  & $300 \times 70$ & 0.5875(34) & 0.6772(116) & 0.8248(115) \\
\hline
\end{tabular}
\caption{The energies, $E(R)$, of the lightest three flux tube states with 
length $R$ in the sector $q=0$ $(--)$.}
\label{table_mmenergy}
\end{center}
\end{table}

\begin{table}[h]
\begin{center}
\begin{tabular}{|c|c|ccccc|}\hline
$R/a$ & $l_\perp\times l_t /a^2 $ & \multicolumn{5}{|c|}{ $aE(R) \ ; \ $ $q=1$ $(+)$ } \\ \hline
20  & \multirow{18}{*}{$70 \times 70$} & 0.4899(45) & 0.5612(88) & 0.6680(73) & 0.6693(192) & 0.8066(195) \\
\cline{1-1}
\cline{3-7}
25  & & 0.4722(37) & 0.5236(92) & 0.6141(132) & 0.6804(73) & 0.7865(307) \\
\cline{1-1}
\cline{3-7}
30  & & 0.4644(29) & 0.4919(111) & 0.6029(64) & 0.6480(153) & 0.7514(132) \\
\cline{1-1}
\cline{3-7}
\multirow{2}{*}{35}  & & 0.4585(41) & 0.5169(85) & 0.5594(197) & 0.6343(83) & 0.7561(34) \\
& & 0.7460(141) & & & & \\
\cline{1-1}
\cline{3-7}
\multirow{2}{*}{40}  & & 0.4778(37) & 0.4953(95) & 0.5904(106) & 0.6632(94) & 0.6684(45) \\
& & 0.7484(61)  & & & & \\
\cline{1-1}
\cline{3-7}
\multirow{2}{*}{45}  & & 0.4820(34) & 0.5359(51) & 0.6110(73) & 0.6414(69) & 0.6353(95) \\
& & 0.7416(71)  & & & & \\
\cline{1-1}
\cline{3-7}
\multirow{2}{*}{47}  & & 0.4801(32) & 0.5323(65) & 0.6249(42) & 0.6377(44) & 0.6633(53) \\
& & 0.7538(74)  & & & & \\
\cline{1-1}
\cline{3-7}
\multirow{2}{*}{50}  & & 0.4919(29) & 0.5509(61) & 0.6414(48) & 0.6300(35) & 0.6583(131) \\
& & 0.7675(98)  & & & & \\
\cline{1-1}
\cline{3-7}
\multirow{2}{*}{52}  & & 0.4898(41) & 0.5536(75) & 0.6269(80) & 0.6236(78) & 0.6246(110) \\
& & 0.7402(138)  & & & & \\
\cline{1-1}
\cline{3-7}
\multirow{2}{*}{54}  & & 0.4996(41) & 0.5856(62) & 0.6359(64) & 0.6267(51) & 0.6441(232) \\
& & 0.7397(84)  & & & & \\
\cline{1-1}
\cline{3-7}
\multirow{2}{*}{55}  & & 0.4938(35) & 0.5478(98) & 0.6345(95) & 0.6167(105) & 0.6546(156)* \\
& & 0.7532(84) & & & &  \\
\cline{1-1}
\cline{3-7}
56  & & 0.5016(51) & 0.5732(75) & 0.6255(70) & 0.6550(48) & 0.6540(58)* \\
\cline{1-1}
\cline{3-7}
\multirow{2}{*}{58}  & & 0.5071(37) & 0.5524(118) & 0.6527(44) & 0.6361(57) & 0.6942(82) \\
& & 0.7587(69)  & & & & \\
\cline{1-1}
\cline{3-7}
60  & & 0.5228(36) & 0.5882(79) & 0.6614(143) & 0.6439(66) & 0.6799(72)* \\ 
\cline{1-1}
\cline{3-7}
65  & & 0.5337(57) & 0.5905(93) & 0.6771(77) & 0.6370(103)* & \\ 
\cline{1-1}
\cline{3-7}
70  & & 0.5452(44) & 0.6166(94) & 0.6734(97)* & & \\ 
\cline{1-1}
\cline{3-7}
75  & & 0.5595(72) & 0.6462(121) & 0.7031(59)* & & \\
\cline{1-1}
\cline{3-7}
80  & & 0.5865(47) & 0.6628(157) & 0.6883(73)* & &  \\ 
\hline
\multirow{2}{*}{40}  & \multirow{2}{*}{$55 \times 55$} & 0.4621(31) & 0.5323(61) & 0.6214(79) & 0.6969(58) & 0.6851(124) \\
& & 0.7858(88) & & & &  \\
\hline
\multirow{2}{*}{40}  & \multirow{2}{*}{$55 \times 70$} & 0.4615(53) & 0.5312(56) & 0.6243(29) & 0.6811(99) & 0.6838(75) \\
& & 0.7738(57)  & & & & \\ 
\hline
\multirow{2}{*}{40}  & \multirow{2}{*}{$65 \times 70$} & 0.4724(25) & 0.5049(53) & 0.6065(52) & 0.6560(93) & 0.6999(53) \\
& & 0.7679(71)  & & & & \\
\hline
\multirow{2}{*}{40}  & \multirow{2}{*}{$80 \times 70$} & 0.4745(33) & 0.5076(48) & 0.5712(97) & 0.5977(127) & 0.6489(77) \\
& & 0.6982(99)  & & & & \\
\hline
40  & $160 \times 70$ & 0.4737(27) & 0.5317(77) & 0.6076(64) & 0.5532(106)* & 0.6793(85) \\
\hline
\multirow{2}{*}{40}  & \multirow{2}{*}{$300 \times 70$} & 0.4701(26) & 0.5394(46) & 0.6040(62) & 0.5587(79)* & 0.6958(42) \\
& & 0.6950(121)  & & & & \\
\hline
\end{tabular}
\caption{The energies, $E(R)$, of the lightest six flux tube states with 
length $R$ in the sector $q=1$ $(+)$.}
\label{table_p=1penergy}
\end{center}
\end{table}

\clearpage

{\footnotesize
\begin{center}
\begin{longtable}{|c|c|ccccc|}

\hline
$R/a$ & $l_\perp\times l_t /a^2 $ & \multicolumn{5}{|c|}{ $aE(R) \ ; \ $ $q=1$ $(-)$ } \\
\hline
\endfirsthead
\hline
$R/a$ & $l_\perp\times l_t /a^2 $ & \multicolumn{5}{|c|}{ $aE(R) \ ; \ $ $q=1$ $(-)$ } \\
\hline
\endhead

\multirow{2}{*}{20}  & \multirow{18}{*}{$70 \times 70$} & 0.4025(14) & 0.5260(73) & 0.6537(90) & 0.6300(52) & 0.6813(202) \\
& & 0.7666(70) & & & & \\
\cline{1-1}
\cline{3-7}
\multirow{2}{*}{25}  & & 0.3621(16) & 0.4946(58) & 0.6232(78) & 0.6326(65) & 0.6182(84) \\
& & 0.7189(66) & & & & \\
\cline{1-1}
\cline{3-7}
\multirow{2}{*}{30}  & & 0.3448(16) & 0.4861(50) & 0.5896(73) & 0.5737(180) & 0.6070(159) \\
& & 0.6968(108) & & & & \\
\cline{1-1}
\cline{3-7}
\multirow{2}{*}{35}  & & 0.3382(19) & 0.4886(37) & 0.5600(89) & 0.6152(71) & 0.5959(140)* \\
& & 0.6828(114) & & & & \\
\cline{1-1}
\cline{3-7}
\multirow{2}{*}{40}  & & 0.3403(14) & 0.4855(52) & 0.5741(107) & 0.6100(70) & 0.6154(54) \\
& & 0.7201(63) & & & & \\
\cline{1-1}
\cline{3-7}
\multirow{2}{*}{45}  & & 0.3467(23) & 0.4857(53) & 0.5562(93) & 0.6007(69) & 0.6247(107) \\
& & 0.6813(228) & & & & \\
\cline{1-1}
\cline{3-7}
\multirow{2}{*}{47}  & & 0.3524(21) & 0.4953(31) & 0.5891(51) & 0.6019(50) & 0.6478(58) \\
& & 0.6982(81) & & & & \\
\cline{1-1}
\cline{3-7}
\multirow{2}{*}{50}  & & 0.3577(24) & 0.4911(59) & 0.5963(74) & 0.5999(59) & 0.6490(71) \\
& & 0.6966(58) & & & & \\
\cline{1-1}
\cline{3-7}
\multirow{2}{*}{52}  & & 0.3665(23) & 0.5012(58) & 0.6073(41) & 0.6179(61) & 0.6777(79) \\
& & 0.6595(143)* & & & & \\
\cline{1-1}
\cline{3-7}
\multirow{2}{*}{54}  & & 0.3700(17) & 0.5075(33) & 0.6171(53) & 0.6229(55) & 0.6671(73) \\
& & 0.6819(52) & & & & \\
\cline{1-1}
\cline{3-7}
\multirow{2}{*}{55}  & & 0.3681(35) & 0.5033(59) & 0.6040(70) & 0.6162(47) & 0.6852(65) \\
& & 0.7166(110)* & & & & \\
\cline{1-1}
\cline{3-7}
\multirow{2}{*}{56}  & & 0.3741(24) & 0.5116(38) & 0.6230(35)* & 0.6196(68) & 0.6727(58) \\
& & 0.6947(70) & & & & \\
\cline{1-1}
\cline{3-7}
\multirow{2}{*}{58}  & & 0.3840(22) & 0.5163(41) & 0.6030(72) & 0.6296(59) & 0.6812(64) \\
& & 0.6810(68)* & & & & \\
\cline{1-1}
\cline{3-7}
\multirow{2}{*}{60}  & & 0.3899(20) & 0.5221(54) & 0.5940(135) & 0.6431(77) & 0.6850(56) \\
& & 0.6740(113)* & & & & \\ 
\cline{1-1}
\cline{3-7}
\multirow{2}{*}{65}  & & 0.4058(23) & 0.5346(49) & 0.6342(53) & 0.6606(68) & 0.6921(83) \\
& & 0.6810(170)* & & & & \\ 
\cline{1-1}
\cline{3-7}
70  & & 0.4230(26) & 0.5555(52) & 0.6418(222) & 0.6895(86) & 0.7063(61)* \\ 
\cline{1-1}
\cline{3-7}
75  & & 0.4357(42) & 0.5623(97) & 0.6660(86) & 0.7014(84)* & \\
\cline{1-1}
\cline{3-7}
80  & & 0.4572(45) & 0.5701(84) & 0.7048(69) & 0.7088(93)* & \\ 
\hline
\multirow{2}{*}{40}  & \multirow{2}{*}{$55 \times 55$} & 0.3424(17) & 0.4991(64) & 0.6044(67) & 0.6363(98) & 0.6713(104) \\
& & 0.7215(61) & & & & \\
\hline
\multirow{2}{*}{40}  & \multirow{2}{*}{$55 \times 70$} & 0.3429(14) & 0.4994(45) & 0.6061(59) & 0.6152(137) & 0.6743(124) \\
& & 0.6909(95) & & & & \\ 
\hline
\multirow{2}{*}{40}  & \multirow{2}{*}{$65 \times 70$} & 0.3420(19) & 0.4876(47) & 0.5947(123) & 0.6052(27) & 0.6154(93) \\
& & 0.7075(118) & & & & \\
\hline
\multirow{2}{*}{40}  & \multirow{2}{*}{$80 \times 70$} & 0.3375(23) & 0.4890(31) & 0.5483(89) & 0.6091(53) & 0.6414(87) \\
& & 0.7307(51) & & & & \\
\hline
\multirow{2}{*}{40}  & \multirow{2}{*}{$160 \times 70$} & 0.3426(11) & 0.4855(31) & 0.5294(73) & 0.5975(59) & 0.5975(32) \\
& & 0.7095(98) & & & & \\
\hline
\multirow{2}{*}{40}  & \multirow{2}{*}{$300 \times 70$} & 0.3375(17) & 0.4843(34) & 0.5386(47) & 0.5863(45) & 0.5760(66) \\
& & 0.7030(81) & & & & \\
\hline

\caption{The energies, $E(R)$, of the lightest six flux tube states with 
length $R$ in the sector $q=1$ $(-)$.} 
\label{table_p=1menergy}
\end{longtable}

\end{center} }

\end{appendix}

\newpage
\bibliographystyle{utphys}
\bibliography{references}

\clearpage

\end{document}